\documentclass[pra,twocolumn,preprintnumbers,amsmath,amssymb,superscriptaddress]{revtex4}

\usepackage{graphicx}% Include figure files
\usepackage{dcolumn}% Align table columns on decimal point
\usepackage{bm}% bold math
\usepackage{braket}
\usepackage{amsmath}
\usepackage{tikz}
\usepackage{cancel}
\usepackage{hyperref}

\begin{document}

\title{Exponential convergence dynamics in Grover's search algorithm}

\author{Samuel Cogan}
\affiliation{New York University Shanghai, NYU-ECNU Institute of Physics at NYU Shanghai, 567 West Yangsi Road, Shanghai, 200126, China.}
\affiliation{Ecole Normale Supérieure Paris-Saclay, 4 Av. des Sciences, 91190 Gif-sur-Yvette, France}

\author{Jonathan Raghoonanan}
\affiliation{New York University Shanghai, NYU-ECNU Institute of Physics at NYU Shanghai, 567 West Yangsi Road, Shanghai, 200126, China.}
\affiliation{Department of Physics, New York University, New York, NY 10003, USA}

\author{Tim Byrnes}
\email[]{tim.byrnes@nyu.edu}
\affiliation{New York University Shanghai, NYU-ECNU Institute of Physics at NYU Shanghai, 567 West Yangsi Road, Shanghai, 200126, China.}
\affiliation{State Key Laboratory of Precision Spectroscopy, School of Physical and Material Sciences, East China Normal University, Shanghai 200062, China}
\affiliation{Center for Quantum and Topological Systems (CQTS), NYUAD Research Institute, New York University Abu Dhabi, UAE.}
\affiliation{Department of Physics, New York University, New York, NY 10003, USA}

\date{\today}% It is always \today, today,

\begin{abstract}
Grover's search algorithm is the cornerstone of many applications of quantum computing, providing a quadratic speed-up over classical methods. One limitation of the algorithm is that it requires knowledge of the number of solutions to obtain an optimal success probability, due to the oscillatory dynamics between the initial and solution states (the ``souffl{\'e} problem''). While various methods have been proposed to solve this problem, each has its drawbacks in terms of inefficiency or sensitivity to control errors. Here, we modify Grover's algorithm so that, for suitably chosen parameters, the usual oscillatory dynamics are replaced by an approximately exponential convergence into the solution subspace. The basic idea is to couple the solution states to an engineered ancilla reservoir such that the initial state is nonreflectively absorbed. Trotterizing the continuous algorithm yields a quantum circuit that gives equivalent performance, while preserving the same quadratic quantum speedup as the original algorithm. 
\end{abstract}

\maketitle

\section{Introduction}
% - 1 - Grover's algorithm
% other applications of grover's + reference
Grover's search algorithm  is one of the central algorithms in quantum computing that provides a provable speed-up compared to classical computing \cite{grover1996fast}. Given an oracle function that recognizes marked items, the task is to locate one of those $ M $ marked states in an unsorted list of $ N $ elements.  The time complexity of standard Grover's algorithm is $ O(\sqrt{N/M}) $, which is asymptotically optimal \cite{bennett1997strengths} and has a quadratic speedup over classical algorithms.  The generic problem setting means that it has a high degree of applicability.  For example, it has found applications in cryptography \cite{Hsu03,Hao2010,tessler2017bitcoin,Sarah2024PracticalGroverAES}, 
matrix and graph problems \cite{Magniez07,5954250}, signal processing and quantum control tasks \cite{Gao2022GWGrover,Chen05}, optimization \cite{Durr96aquantum, Grover1997,Campbell2019CSP,Bennett1997,Furer2008,Morimoto2024QuADS}, element distinctness \cite{2005quant.ph.4012A}, collision problems \cite{Brassard1997}, and quantum machine learning \cite{Aimeur2013,liao2024quadratic,Morales_2018}. 

\begin{figure}[t]
\includegraphics[width=\columnwidth]{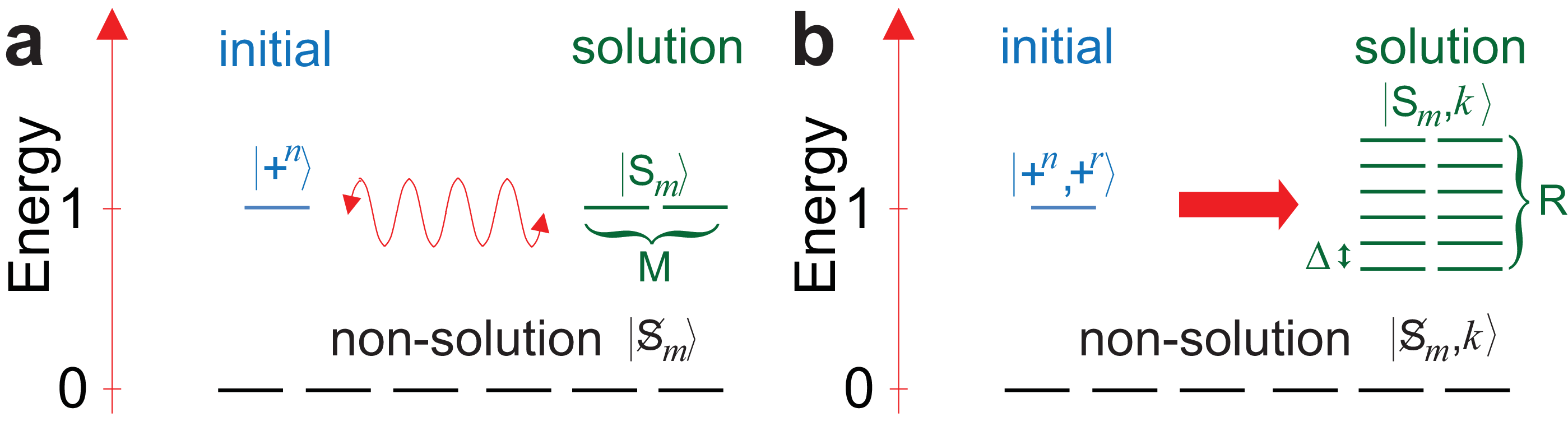}
\caption{
Grover's algorithm in continuous Hamiltonian formulation.   (a) Standard Grover's algorithm and (b) dissipative Grover's algorithm proposed in this work.  Examples shown are for $ M = 2$ solutions for both cases.  
\label{fig1}
}
\end{figure}

% - 2 - Standard grover, how it works
% - 3 - problem: nb of targets states
It is well known that the evolution of Grover's algorithm can be considered a rotation in a two-dimensional space, spanned by the initial state (typically taken to be an equal superposition state of computational basis states) and the superposition of solution states \cite{nielsen2010quantum}.  While Grover's algorithm is typically formulated as a gate-based quantum algorithm where the resources are evaluated in the number of oracle calls, it can also be considered in a continuous setting, where a fixed Hamiltonian determines its evolution \cite{byrnes2018generalized,nielsen2010quantum}.  In this formulation, the initial state and the target states are energetically separated from the remaining states, and the time evolution corresponds to Rabi oscillations (see Fig. \ref{fig1}(a)).  This oscillatory nature of Grover's algorithm is a weakness of the algorithm (the ``soufflé problem'' \cite{yoder2014fixed}) since the frequency of the oscillations depends on the number of solutions $ M $.  The number of Grover iterates that should be applied (or the time the Hamiltonian is applied) such that a high success probability is obtained then also depends upon $ M $,  which may not be known in advance.

% - 4 - solutions that already exists

Several approaches have been developed to overcome this limitation. The first solution to this problem is quantum counting, which performs phase estimation on the Grover iterate to estimate the effective rotation angle and thereby infer $M$ before running a standard Grover search~\cite{nielsen2010quantum}. Another route is to design fixed-point quantum search algorithms, in which the success probability is monotonically non-decreasing with the number of queries. Examples include Grover's $\pi/3$ algorithm~\cite{grover2005fixed,grover2006quantum}, which recursively replaces the usual $\pi$ phase inversions by smaller phase shifts; Long's phase–matching variant with zero theoretical failure rate~\cite{Long_2001}; and more recent fixed-point constructions derived from the quantum singular value transformation (QSVT) or quantum signal processing framework~\cite{yoder2014fixed,martyn2021grand}. In these QSVT-based approaches, one engineers a polynomial filter in the Grover iterate through a carefully chosen sequence of single-qubit phase rotations; this yields a fixed-point search that retains the optimal $O(\sqrt{N/M})$ asymptotic scaling up to logarithmic factors in the target error. \textcolor{black}{Related Grover-inspired variational ans\"atze have also recently been proposed as another route to search dynamics with quadratic scaling and reduced dependence on the marked-state degeneracy~\cite{Zhang2024,XieGroverMixer2025}.}

However, these approaches come with trade-offs. Quantum counting requires running a separate phase-estimation subroutine prior to the search, which increases depth and circuit complexity. Fixed-point and exact amplitude-amplification methods rely on precisely calibrated phase rotations, which are susceptible to control errors. Currently, there is no compact and robust way to solve the souffl{\'e} problem for quantum search.

% - 6 - presentation of our scheme
In this paper, we present a variant of Grover's algorithm (which we call {\it dissipative Grover's algorithm}) that \textcolor{black}{converts the oscillatory dynamics of the original algorithm into an approximately exponential decay up to finite-reservoir revivals}. The basic idea is best understood in the continuous Hamiltonian formulation. By adding extra ancilla qubits, we split the energy spectrum of the solution states into a spread of energies, forming a  reservoir (see Fig. \ref{fig1}(b)).  In standard Grover's algorithm, the system undergoes Rabi oscillations because there are only two levels involved.  With our modification, the initial state dissipatively decays into the solution states. Due to the exponential decay dynamics, the algorithm becomes far less sensitive to the evolution time, which \textcolor{black}{removes the need to know the number of solutions $M$ in order to set a sharply defined stopping time} and is robust under control errors.

\section{Continuous time formulation}

Let us first start with the continuous time formulation of standard Grover's algorithm.  Consider initially preparing the register of the quantum computer consisting of $n $ qubits in the state $ | \psi_0 \rangle = |+^n \rangle := |+ \rangle^{\otimes n} $, where $ | + \rangle = (|0 \rangle + | 1 \rangle)/\sqrt{2} $. The dimension of the search space in this case is $ N = 2^n $.  Then, apply the Hamiltonian 
\begin{align}
H_\text{G} = |+^n\rangle\langle +^n| + \sum_{m=0}^{M-1} |S_m\rangle\langle S_m|
\label{standardgroverham}
\end{align}
where $\ket{S_m} $ is one of the $ M $ solution states in the computational basis.  The Hamiltonian (\ref{standardgroverham}) can be interpreted as specifying the energies of the states $ |+^n \rangle, | S_m \rangle $  to be 1, while all other states being energy zero (Fig. \ref{fig1}(a)).  Standard analysis yields an oscillation of the amplitude between the initial state and $\sum_m |S_m \rangle/\sqrt{M}  $ with period $ \propto \sqrt{N/M} $ (see Appendix \textcolor{black}{\ref{app:standard}}) \cite{nielsen2010quantum,byrnes2018generalized}.   \textcolor{black}{We assume the standard Grover scenario where it is known how to construct $ H_G $ (or equivalently the oracle in the discrete-time case), but neither the solution states $ |S_m \rangle $ or the number of solutions $ M $ is known \cite{nielsen2010quantum}.} 

To suppress the oscillatory dynamics, we propose modifying the Hamiltonian according to 
\begin{align}
H_\text{DG} = & |+^n\rangle\langle +^n| \otimes |+^r\rangle\langle +^r|  \nonumber \\
& + \sum_{m=0}^{M-1} |S_m\rangle\langle S_m|  \otimes \sum_{k=0}^{R-1}E_k |k\rangle\langle k| .
\label{mainham}
\end{align} 
Here we have added $ r $ ancilla reservoir qubits and have defined $ R = 2^r$. The states $ | k \rangle $ are computational basis states with the binary decomposition of the integer $ k $. In this paper, we use $ E_k = 1  + \Delta (k-R/2+1/2) $ throughout, which gives a ladder of states spaced by $ \Delta $. \textcolor{black}{The number of reservoir states is therefore $R=2^r$. The parameters $ R $ and $ \Delta $ are tunable algorithmic parameters.  Increasing $R$ makes the ancilla states more closely approximate an infinite reservoir such that exponential decay dynamics occur. }

\begin{figure}[t]
\includegraphics[width=\columnwidth]{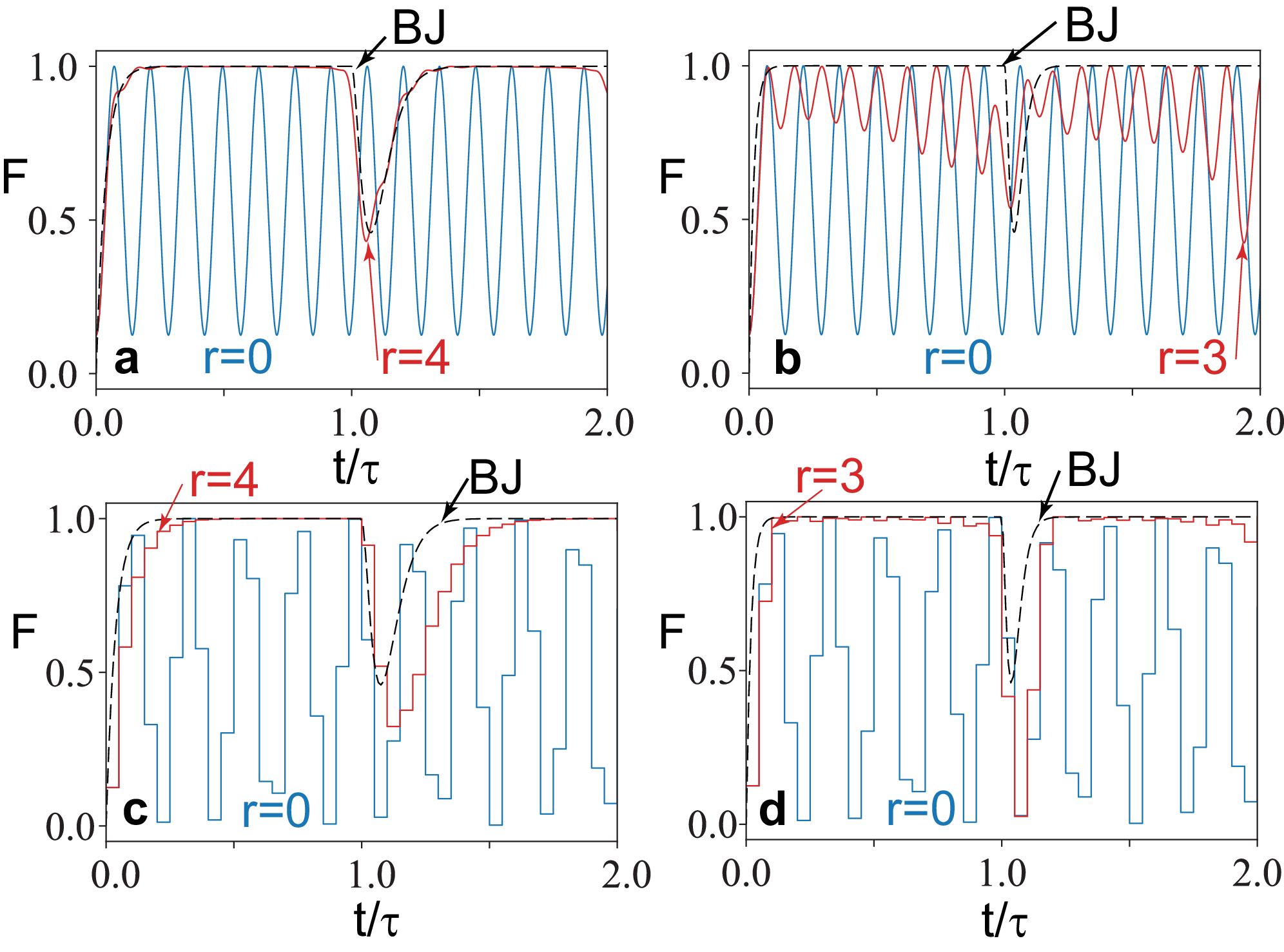}
\caption{Success probability (\ref{fidelity}) of obtaining a solution state under various versions of Grover's algorithm.  Standard Grover's algorithm corresponds to $ r = 0 $, and any of the values with $ r > 0 $ correspond to dissipative Grover's algorithm.  In (a)(b), the $ r = 0 $ line is obtained by time evolving the Hamiltonian (\ref{standardgroverham}) from the initial state $ | \psi_0 \rangle $, while the $ r = 3,4$ line is obtained using Hamiltonian (\ref{mainham}). In (c)(d), the $ r = 0 $ line is obtained by standard Grover's algorithm, while the $ r = 3,4 $ line is obtained using 
 (\ref{trottered}) with $ \delta t = \pi $. Dashed lines are the  fidelity curves  (\ref{fidbj2}) for the BJ model.  All figures were obtained with $n=3$, $M=1 $, $ \Delta = 0.1$.  
\label{fig2}
}
\end{figure}

\begin{figure}[t]
\includegraphics[width=0.95\columnwidth]{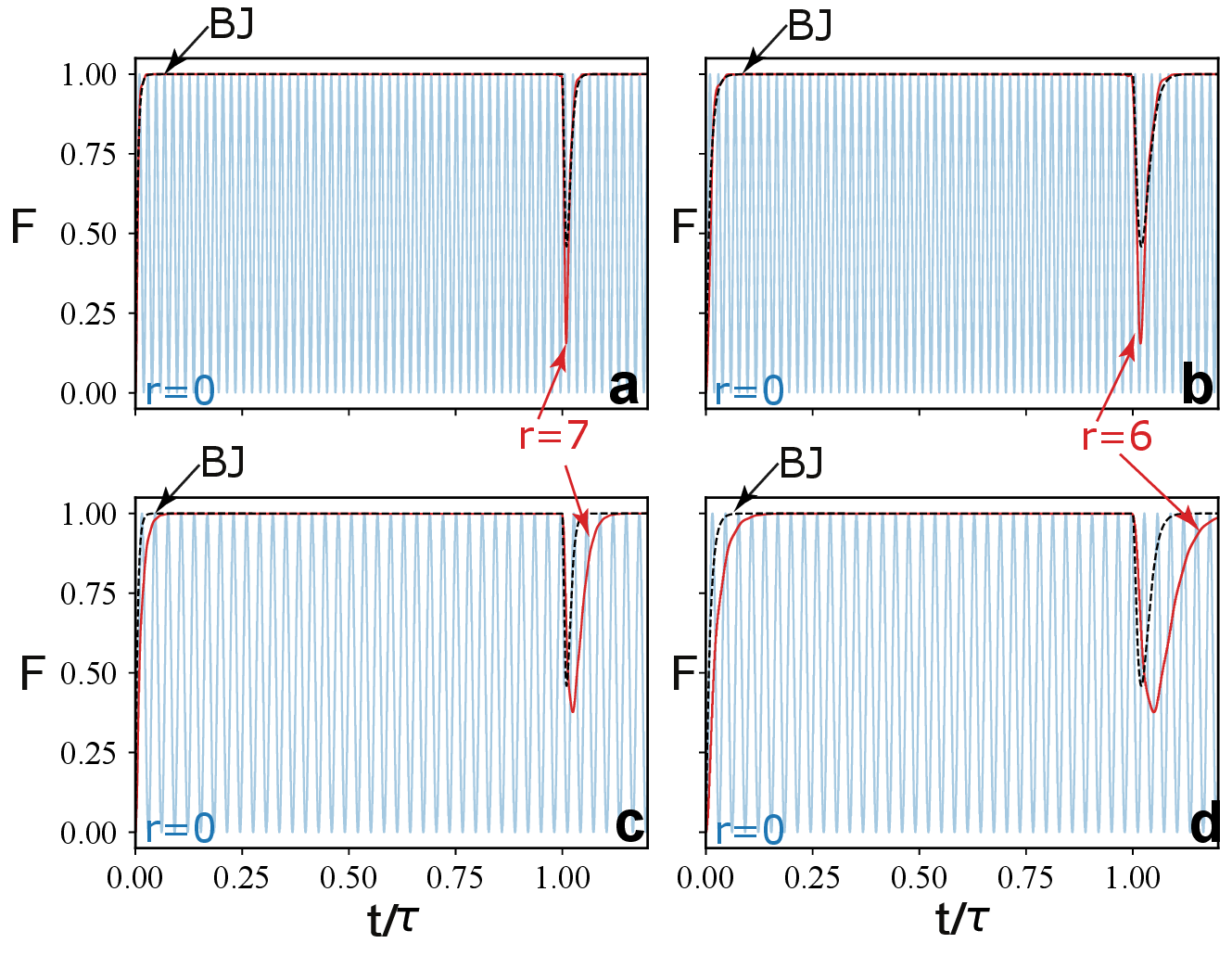}
\caption{\textcolor{black}{Success probability (\ref{fidelity}) of obtaining a solution state under various versions of Grover's algorithm with larger register sizes. The search register has $n=10$ qubits with $M=1$ marked state and $\Delta$ is set according to (\ref{rdeltachoice}) using $C=5$. (a)(b) Continuous-time evolution under the Hamiltonian formulation. (c)(d) Corresponding discrete-time Grover-like product-formula implementation with $\delta t=\pi$. $r=0$ curves correspond to standard Grover search without a reservoir register, while $ r = 6,7 $ correspond to dissipative Grover search. The dashed  curves show the BJ reservoir approximation.  
}
}
\label{fig3}
\end{figure}

The Hamiltonian is then evolved in time, starting with the state $ | \psi_0 \rangle = |+^n , +^r \rangle := |+^n \rangle \otimes |  +^r \rangle $. Some representative time evolutions are shown in Fig. \ref{fig2}(a) and Fig. \ref{fig3}(a)(b).  Shown is the probability of obtaining the solution state 
%
% \textcolor{blue}{
\begin{align}
    F(t)
    &= \sum_{m=0}^{M-1}
    \langle \psi(t)| {\bigl( |S_m\rangle\langle S_m|\otimes I \bigr)} |\psi(t) \rangle \nonumber\\
    &= \sum_{m=0}^{M-1}\sum_{k=0}^{R-1}
    \left| \langle S_m,k | \psi(t) \rangle\right|^2 
    \label{fidelity}
\end{align}
% }
%
where $  | \psi (t) \rangle = e^{-iHt} | \psi_0 \rangle $ \textcolor{black}{and $I $ is the identity matrix. For $r=0$, $R=1$ this reduces to the usual Grover success probability.}
The probability quickly converges to values near $1$.  For comparison, we also show the corresponding evolution for the standard Grover Hamiltonian (\ref{standardgroverham}).  We see that, as expected, the fidelity undergoes oscillations, such that a precise time is required to obtain a high fidelity.  At multiples of particular times $ \tau $, the dissipative Grover dynamics exhibit revivals due to the finite size of the reservoir.  Apart from these times, the fidelity remains near unity.  For an insufficient number of reservoir states the dynamics deviates from the exponential evolution (Fig. \ref{fig2}(b)).

\section{Scaling and revival times}

% {\color{blue} 
\subsection{Bixon-Jortner model}
% }

The numerics shown in Fig. \ref{fig2}(a) confirm the expected behavior of the dissipative Grover evolution.  We now obtain analytical estimates of the performance.  Our approach will be to approximate the Hamiltonian (\ref{mainham}) as having similar dynamics as the Bixon-Jortner (BJ) model \cite{bixon1968intramolecular}, which possesses analytical solutions.  The BJ model is a model of a single source state decaying to a reservoir consisting of an evenly spaced infinite ladder of orthogonal states (see Appendix \textcolor{black}{\ref{app:bjmodel}}).   To make an equivalence between Hamiltonian (\ref{mainham}) and the BJ model, 
we follow a similar procedure to the analysis of Eq. (\ref{standardgroverham}) and  orthogonalize the states $ | +^n, +^r \rangle $ and $ | S_m, k \rangle $ (see Appendix \textcolor{black}{\ref{app:mapping}}). We decompose $\ket{+^n } = \sqrt{\frac{M}{N}}\ket{\tilde{S}_0 } + \sqrt{\frac{N-M}{N}}\ket{\perp^n }$, where $ \ket{\tilde{S}_p } = \sum_{m=0}^{M-1} e^{2 \pi i p m/M}  | S_m \rangle/\sqrt{M}  $ and $ \ket{\perp^n } = \sum_{m=0}^{N-M-1} | \cancel{S}_m \rangle/\sqrt{N-M}  $, where $ | \cancel{S}_m \rangle $ labels the complement of the solution states \textcolor{black}{(i.e. non-solution states)} such that 
$\langle \tilde{S}_p  | \perp^n \rangle = 0 $.  The Hamiltonian (\ref{mainham}) is then written in this basis as
\begin{align}
& H_\text{DG} =  \sum_{k=0}^{R-1}(E_k+\frac{M}{NR}) |\tilde{S}_0,k \rangle\langle \tilde{S}_0,k| \nonumber \\ 
& + (1 - \frac{M}{N}) |\perp^n ,+^r \rangle\langle \perp^n ,+^r |  \nonumber \\
& + \frac{1}{N}\sqrt{\frac{M(N-M)}{R}}  \sum_{k=0}^{R-1} ( 
|\tilde{S}_0,k \rangle\langle \perp^n ,+^r | + \text{H.c.})\nonumber \\
& + \frac{M}{NR} \sum_{k \ne k'} |\tilde{S}_0,k \rangle\langle \tilde{S}_0,k'|
+ \sum_{p=1}^{M-1} \sum_{k=0}^{R-1}E_k |\tilde{S}_p,k \rangle\langle \tilde{S}_p,k|,
\label{bjlikeham}
\end{align}
where we used $ | +^r \rangle = \sum_{k=0}^{R-1} | k \rangle/\sqrt{R} $. This Hamiltonian takes a form that is very similar to the BJ model if we identify $\ket{\perp, +^r}$ as the source state and the $\ket{\tilde{S}_0, k}$ as the reservoir states. The differences to the BJ model are: (i) that the number of reservoir states $ R $ is finite; (ii) there is an additional term (the second last term in (\ref{bjlikeham})) that involves off-diagonal terms between the states $ |\tilde{S}_0, k \rangle $.  There is also an additional term that involves the $ \ket{\tilde{S}_p }  $ states for $p \in [1,M-1] $ but similarly to the standard Grover Hamiltonian, this plays no role in the dynamics if the initial state is taken as $ | +^n, +^r \rangle $, due to the block diagonal nature of the Hamiltonian.

Assuming that $ R \gg 1 $ and $ M \ll N $ such that the differences (i) and (ii) can be suitably neglected, we may use the exact solution of the BJ model to obtain the evolution of the fidelity (\ref{fidelity}) for times $ 0 \le t < 2\tau $ \textcolor{black}{(see Appendix \ref{app:mapping})}
%
% \textcolor{blue}{
\begin{align}
F_{\rm BJ}(t) &=
1-\frac{N-M}{N}\left|\tilde{a} (t)\right|^2 ,
\label{fidbj2}\\
\tilde{a}(t) &=
e^{-\gamma t/2} - \gamma e^{-\gamma(t-\tau)/2} (t-\tau)\Theta(t-\tau).
\label{abj}
\end{align}
% }
%
where $ \Theta(t) $ is the Heaviside step function and
\begin{align}
\gamma = 2\pi\frac{M(N-M)}{R \Delta N^2 } .
\label{gammabj}
\end{align}
The revival time is given by
\begin{align}
    \tau = \frac{2\pi}{\Delta} .
    \label{taubj}
\end{align}
In Fig. \ref{fig2}(a)(b) we compare the dynamics predicted by the BJ model with the exact dynamics. The BJ model captures the main qualitative features of the dynamics, in particular the initial decay scale and the approximate revival time. Deviations are expected for the small reservoirs used in Fig.~\ref{fig2}, because the reservoir is finite and the additional off-diagonal term in (\ref{bjlikeham}) is not completely negligible. As $R$ is increased,  the dynamics approach the BJ prediction more closely.

\textcolor{black}{
Figure~\ref{fig3}(a)(b) shows results for a larger search space with $n=10$ and a larger reservoir $ r = 6,7 $. Compared with the small-system example in Fig.~\ref{fig2}, the dissipative curves are smoother over the pre-revival window and remain close to unity after the initial transfer into the marked subspace. The revival times are much longer than the typical decay time  $ \tau \gg 1/\gamma $, resulting in high fidelities.    The larger reservoir acts to reduce the residual oscillations between the revival times by suppressing reflections.   This is also consistent with the BJ interpretation: increasing the data Hilbert-space dimension reduces the initial marked-state overlap $M/N$ and produces a clearer separation between the source-like component and the reservoir-like marked manifold.  
The finite-reservoir revival near $t \approx  \tau$ is still present, as expected from the finite ladder spacing, but away from the revival, the dissipative dynamics exhibit the same approximately exponential convergence and broad high-fidelity plateau observed in the smaller system.
}

% {\color{blue} 
\subsection{Optimal parameter choices}
% }

The parameters $ N, M $ are fixed by the search problem to be solved, hence there are two free parameters $ R, \Delta $ which may be chosen to best observe the exponential dynamics.  We now discuss the best choices of the free parameters.  

First, we would like that the time of the revival is sufficiently separated from the decay dynamics taking place on the timescale such that  $ 1/\gamma \ll \tau  $.  This allows a high fidelity to be obtained $F \approx 1 - e^{- \gamma \tau }$ by choosing a time just before the revival time. Substituting the expressions (\ref{gammabj}) and (\ref{taubj}), this criterion leads to Condition 1 requiring that
\begin{align}
R \Delta^2 \ll 4 \pi^2 \frac{M(N-M)}{N^2} . 
\label{crit1}
\end{align}
Second, as seen in Fig. \ref{fig2}(b), for improperly chosen parameters, the exponential decay gives additional oscillations, which can be estimated to have an amplitude  (see Appendix \textcolor{black}{\ref{app:bjmodelfinite}})
\begin{align}
    \Gamma  = \frac{M(N-M)}{(R N \Delta)^2}  .
\end{align}
Setting this to be $ \Gamma  \ll 1 $ gives the Condition 2:
\begin{align}
\frac{M(N-M)}{N^2} \ll R^2 \Delta^2 . 
\label{crit2}
\end{align} 
Using (\ref{crit2}), we then make the choice
\begin{align}
\Delta = C \frac{\sqrt{M(N-M)}}{N R }
\label{rdeltachoice}
\end{align}
where $ C $ is a constant such that $ \Gamma  \ll 1 $ (we find that in practice $ C > 5 $ suppresses additional oscillations).  Substituting this into Condition 1 leads us to the equivalent constraint that 
\begin{align}
    C^2 \ll 4 \pi^2 R = 4 \pi^2 2^r ,
    \label{rchoice}
\end{align}
which is easily satisfied by taking sufficiently many qubits in the ancilla register. \textcolor{black}{Increasing $r$ improves the continuum-reservoir approximation with dynamics approaching an exponential decay.} The two choices (\ref{rdeltachoice}) and (\ref{rchoice}) fix the free parameters $ R, \Delta $.  

Using the optimized choices we may estimate the time scaling of the dissipative Grover algorithm as
\begin{align}
T \approx \frac{1}{\gamma} = \frac{C}{2 \pi }\frac{N}{\sqrt{M(N-M)}} .
\label{scalings}
\end{align}
This gives the same scaling $ O(\sqrt{N/M}) $ as standard Grover's algorithm up to prefactors, assuming $ M \ll N $, as expected from optimality arguments \cite{bennett1997strengths}. There is a moderate prefactor overhead arising from modifying the dynamics to an exponential decay.

% {\color{blue}
\section{Unknown number of solutions}
\label{sec:unknown}

One of the potential applications of dissipative Grover's algorithm is the case where the number of marked states $M$ is not known in advance.   The choice (\ref{rdeltachoice}) requires knowledge of $ M $ and raises the question of how $ R, \Delta $ should be chosen in this case.  In this section, we give an alternative choice of these parameters suitable for when $ M $ is unknown.  
% As in standard Grover search, we assume access to a marking oracle, or to its Hamiltonian analogue, which distinguishes marked from unmarked basis states. The unknown quantity is not the oracle itself, but the degeneracy $M$ of the marked subspace.  In standard Grover search this degeneracy fixes the rotation angle and hence the optimal stopping time.  In the dissipative construction it fixes the decay rate, but the success probability remains large over a broad time window after the initial decay and before the first finite-reservoir revival.

In this case, we start from Condition 1, Eq. (\ref{crit1}), and use the fact that for any $ M > 1 $, $ 1/N < M(N-M)/N^2 $.  Thus if we demand that 
\begin{align}
    R \Delta^2 \ll 4 \pi^2 \frac{1}{N} ,
    \label{cond1unknownm}
\end{align}
it then follows that Condition 1, Eq. (\ref{crit1}), will also be satisfied for any $ M $.  We may then make the choice
\begin{align}
\Delta = \frac{2\pi}{\sqrt{CNR}},
\label{deltaunknownchoice}
\end{align}
where $C > 1 $ is a constant chosen sufficiently large such that (\ref{cond1unknownm}) is satisfied.  

Substituting this choice into the Condition 2, Eq.  (\ref{crit2}), we require
\begin{align}
  \frac{M(N-M)}{N^2} \ll \frac{4 \pi^2 R}{CN}   
  \label{cond2unknown}.
\end{align}
Using the fact that $ N-M < N $ for any $ M> 1 $, we then demand that
\begin{equation}
R\gg \frac{C M}{4\pi^2}.
\label{rgtcm}
\end{equation}
With this, it follows that (\ref{cond2unknown}) is also satisfied, and hence Condition 2 is satisfied for any $ M > 1 $.  

Strictly speaking, to satisfy (\ref{rgtcm}) we still require some knowledge of the number of solutions $M $.  However, the required precision of this knowledge is rather low due to the $ \gg $ condition.  Thus, the knowledge of the order of magnitude of  $M $  would be sufficient. Alternatively,  $ R $ could be set conservatively large.  

Applying (\ref{deltaunknownchoice}), the decay time is
\begin{align}
T
\approx
\frac{1}{\gamma}
&=
% \textcolor{blue}{
\frac{N^2}{M(N-M)}
\sqrt{\frac{R}{CN}}
% }  
\nonumber\\
&\approx
\frac{1}{M}\sqrt{\frac{RN}{C}},
\qquad (M\ll N) .
\end{align}
We thus see that for fixed $ R, C $, the solution time scales proportional to $ \sqrt{N} $.  Thus with these parameters fixed, the standard Grover scaling is recovered.  On first glance, it appears to have a better scaling with $ M $, however, due to the constraint (\ref{rgtcm}), $ R/C $ is at least of the magnitude of $ M $, hence there is no additional improvement over the standard Grover case.  If $ R $ is set conservatively large, then this will add an additional overhead.  Thus, there is a trade-off of satisfying (\ref{rgtcm}) and the decay time.  We note that for the choice $ R = MC$ the time scaling reverts to the optimal solution.  

% For sparse search problems with $M=O(1)$,  this means that $r=\log_2R$ can be treated as a constant overhead.  If one wants uniform performance for larger possible marked fractions, then $R$ must be increased accordingly, which increases the runtime prefactor. 

\begin{figure}[t]
    \centering
    \includegraphics[width=\columnwidth]{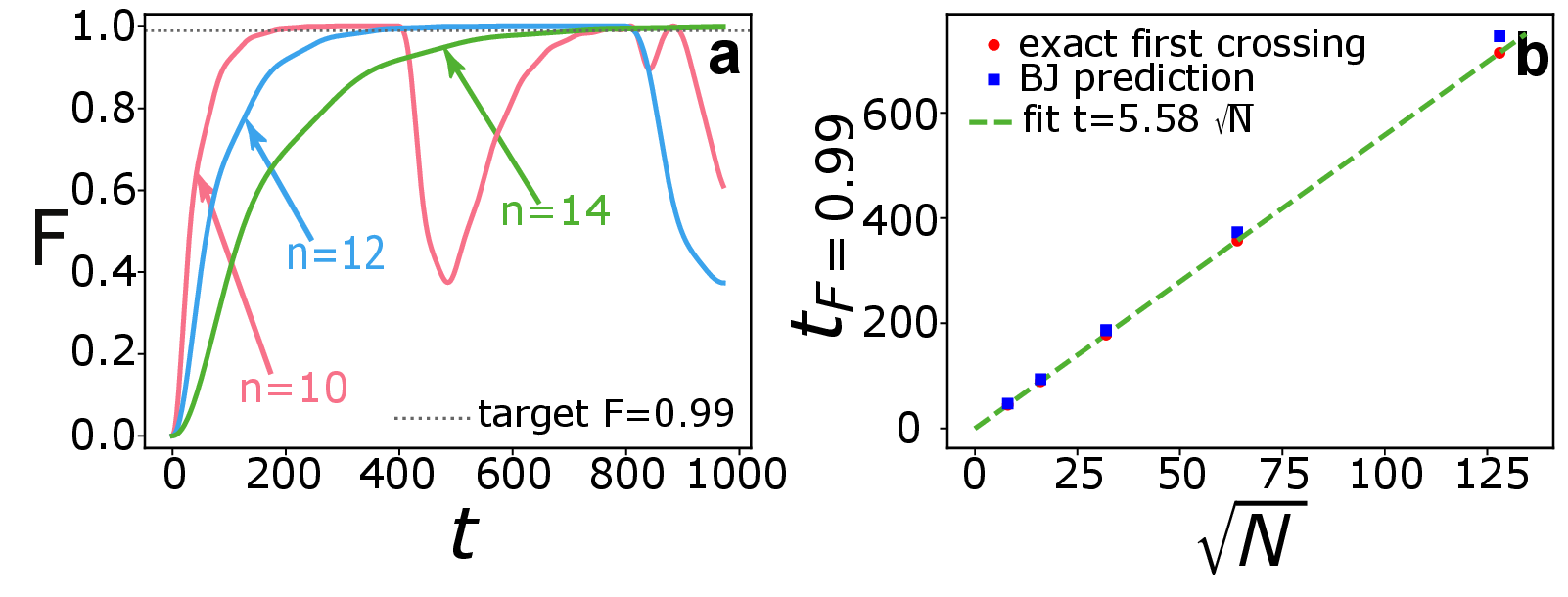}
    \caption{\textcolor{black}{
    Numerical scaling test for the unknown-$M$ parameter choice in the worst sparse case $M=1$.  We use the conservative level spacing (\ref{deltaunknownchoice}), with fixed $r=4$, $R=16$, and $C=10$. Panel (a) shows the exact success probability $F(t)$, for $n=10,12,14$. The dotted horizontal line marks the target $F=0.99$.  Panel (b) shows the extracted first-crossing time $t_{F=0.99}$ as a function of $\sqrt{N}$.  The approximately linear dependence confirms the predicted $O(\sqrt{N})$ worst-case scaling for fixed reservoir size and fixed $C$.  The Bixon--Jortner prediction is shown for comparison; the dashed line is a least-squares fit to the exact numerical crossing times.}
    }
    \label{fig:unknownM-scaling}
\end{figure}

Figure~\ref{fig:unknownM-scaling} verifies this scaling numerically for the worst case $M=1$.  The simulations use the conservative unknown-$M$ choice (\ref{deltaunknownchoice}), while  varying the search register size $N=2^n$ to observe the $ N $ scaling. Figure \ref{fig:unknownM-scaling}(a) shows the success probability of the target state versus time.  We see that the success probability reaches the target $F=0.99$ before the first reservoir revival for each value of $N$.  Figure \ref{fig:unknownM-scaling}(b) extracts the first-crossing time $t_{F=0.99}$.  A fit of the exact numerical crossing times gives $t_{F=0.99} \approx  5.6\sqrt{N}$ for the parameters shown.  The BJ estimate gives the same asymptotic scaling of $t_{F=0.99}^{\rm BJ} \approx  5.8 \sqrt{N}$ (see Appendix \ref{app:unknown}).  
Thus the numerical data confirm that, for fixed $R$ and $C$, choosing the first-crossing time according to the slowest sparse case gives the expected $O(\sqrt N)$ behavior without requiring prior knowledge of $M$. Increasing $R$ improves the smoothness of the finite reservoir and suppresses residual oscillations, but it also increases the runtime prefactor by $\sqrt R$.

% }

\section{Quantum circuit formulation}

\begin{figure}[t]  % Use figure* for full-width
    \centering
    \includegraphics[width=\columnwidth]{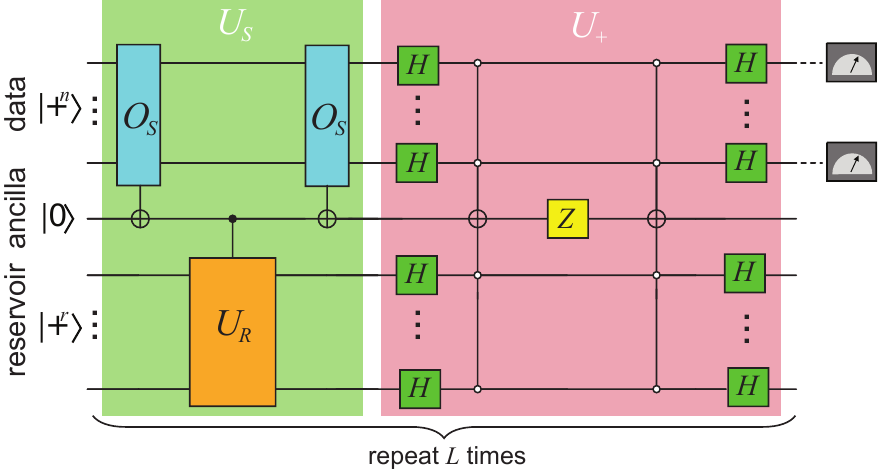}
    \caption{Quantum circuit for the discrete version of the dissipative Grover algorithm (\ref{trottered}). Explicit circuits corresponding to the operators $ U_S $ and $ U_+ $ defined in (\ref{uplusdef}) and (\ref{usdef}) respectively are shown. The oracle is defined as $ O_S = \sum_{m=0}^{M-1} |S_m \rangle \langle S_m | \otimes X
    +  \sum_{m=0}^{N-M-1} | \cancel{S}_m \rangle \langle \cancel{S}_m | \otimes I $, where $ | \cancel{S}_m \rangle  $ are the non-solution states, and the Pauli operator $ X $ acts on the ancilla qubit \cite{nielsen2010quantum, yoder2014fixed}.  
 The conditional reservoir operator is defined as 
    $ U_R =  |1 \rangle \langle 1 | \otimes  \exp( -i \delta t \sum_{k=0}^{R-1}E_k |k \rangle\langle k| )  + |0 \rangle \langle 0 |\otimes  I  $,  where the $ |0 \rangle, | 1 \rangle $ act on the ancilla qubit. $ H $ is the Hadamard operator. 
    }
    \label{fig:quantumcircuit}
\end{figure}

To obtain a quantum circuit corresponding to the dissipative Grover Hamiltonian (\ref{mainham}), we can Trotterize the time evolution as 
\begin{equation}
| \psi(t) \rangle = e^{-iHt}| \psi_0 \rangle  \approx (U_+ U_S)^L | \psi_0 \rangle
\label{trottered}
\end{equation}
where $\delta t = t/L$. Here we defined the Grover iterate operators
\begin{align}
U_+ & = e^{-i |+^n,+^r \rangle\langle +^n,+^r| \delta t} \nonumber \\
    & =I - (1 - e^{-i\delta t}) |+^n,+^r \rangle\langle +^n,+^r|, \label{uplusdef}  \\
U_{S} & = e^{-i \sum_m \sum_k E_k |S_m,k \rangle\langle S_m,k| \delta t}  \nonumber  \\
    & = I - \sum_{m=0}^{M-1} \sum_{k=0}^{R-1}  (1- e^{-i E_k \delta t}) |S_m,k \rangle\langle S_m,k| . \label{usdef} 
\end{align}
To minimize the number of Grover steps $ L $, we choose $ \delta t $ as large as possible while maintaining the integrity of the algorithm. In a similar way to standard Grover's algorithm, we choose $ \delta t = \pi $, which for $ U_+ $ gives a phase inversion operator of the state $ | +^n,+^r \rangle $.  \textcolor{black}{Smaller values of $\delta t$ more closely approximate the continuous Hamiltonian evolution, but require proportionally more steps.}  The operator that corresponds to the oracle is however modified due to the energy spread of the reservoir, where the phase $ e^{-i \pi E_k } $ is applied to the state $ | S_m,k \rangle  $. 

The explicit quantum circuit for our algorithm is shown in Fig. \ref{fig:quantumcircuit}.   After application of the oracle $ O_S $ to identify the solution states by flipping an ancilla, the operator $ U_R $ is applied which puts a phase  $ e^{- i \delta t E_k } $ on the reservoir state  $|k \rangle $ conditionally on the ancilla. An uncompute step disentangles the ancilla qubit and returns it to $ | 0 \rangle $.  A similar procedure is used to construct $ U_+$.  Working in the $ | \pm \rangle $ basis by applying Hadamard gates to all $ n+r $ qubits, the state $ | +^n, +^r \rangle $ can be identified by applying a multi-conditional CNOT gate which flips the ancilla qubit for this state.  By applying a Pauli $ Z $ gate on the ancilla, this creates a phase kickback on $ | +^n, +^r \rangle $.  Uncomputing to disentangle the ancilla qubit returns it to $ | 0 \rangle $, completing the operation.

\begin{table*}[t]
\begin{ruledtabular}
\begin{tabular}{lccc}
Method
& Extra qubits
& Query scaling
& Circuit complexity per Grover iterate \\
\hline
Standard Grover~\cite{grover1996fast,yoder2014fixed}
& none
& $O(\sqrt{N/M})$
& $ O(n^2) $ \\

Quantum counting + Grover~\cite{Brassard98, Brassard_2002}
& $b$
& $  O(2^b) + O(\sqrt{N/M})  $ 
& $ O(n^2) $   \\

Fixed-point/QSVT search~\cite{yoder2014fixed,martyn2021grand}
& none
& $O(\log(1/\epsilon)\sqrt{N/M})$
& $ O(n^2) $  \\

Dissipative Grover (this work)
& $r$
& $O( \sqrt{N/M}  )$
& $ O((n+r)^2) $ \\
\end{tabular}
\end{ruledtabular}
\label{tab:resources}
\caption{\textcolor{black}{
Resource comparison of quantum search variants. Here $N=2^n$, $M$ is the number of marked states, $R=2^r$ is the reservoir dimension, $b$ is the number of qubits in the quantum counting register, and $\epsilon$ denotes a target failure tolerance for fixed-point methods. The circuit complexity per Grover iterate assumes a decomposition of the multi-controlled CNOT into elementary gates, which costs $ O(n^2) $ gates \cite{nielsen2010quantum}.   For quantum counting, a separate algorithm is run separately on the oracle to estimate $ M $, with a query cost of $ O(2^b) $.  For fixed-point methods we assume that the phase of each iterate can be adjusted, hence there is no additional gate complexity of a Grover iterate.   }
}
\end{table*}

% \begin{table*}[t]
% \caption{
% Schematic resource comparison of search variants. Here $N=2^n$, $M$ is the number of marked states, $R=2^r$ is the reservoir dimension, $b$ is the number of qubits in the counting register, and $\epsilon$ denotes a target failure tolerance for fixed-point methods. The query column denotes Grover-like oracle iterations or controlled Grover powers, not elementary one- and two-qubit gates. 
% }
% \begin{ruledtabular}
% \begin{tabular}{lccc}
% Method
% & Extra qubits
% & Query scaling
% & Dominant non-oracle overhead \\
% \hline
% Standard Grover~\cite{grover1996fast,yoder2014fixed}
% & $O(1)$
% & $O(\sqrt{N/M})$
% & reflection about $\ket{+^n}$ \\

% Quantum counting + Grover~\cite{Brassard98, Brassard_2002}
% & $b+O(1)$
% & $O(2^b)+O(\sqrt{N/M})$
% & phase-estimation register and controlled Grover powers \\

% Fixed-point/QSVT search~\cite{yoder2014fixed,martyn2021grand}
% & $O(1)$
% & $O(\sqrt{N/M}\log(1/\epsilon))$
% & calibrated sequence of phase rotations \\

% Dissipative Grover (this work)
% & $r+O(1)$
% & $O(\sqrt{N/M})$
% & reflection about $\ket{+^n,+^r}$ plus controlled reservoir phases \\
% \end{tabular}
% \end{ruledtabular}
% \label{tab:resources}
% \end{table*}

\textcolor{black}{The additional gate overhead in implementing the dissipative Grover iterate is as follows.  Assuming the oracle $ O_S$ is given, implementing $U_S$ requires compute/uncompute of the marked flag around the controlled reservoir phase.  Since the ladder $E_k$ is linear in the reservoir label $k$, this phase can be implemented, up to an overall phase, using $O(r)$ controlled single-qubit $Z$ rotations.  The main overhead is then the reflection $U_+$ about $\ket{+^n,+^r}$, implemented by Hadamard gates on the $n+r$ qubits followed by a multi-conditional CNOT gate.  The multi-conditional CNOT gate can be implemented from elementary gates that grow as $O( (n+r)^2) $ \cite{nielsen2010quantum}. Standard Grover's algorithm requires the same type of gate for phase inversion, which requires $O( n^2) $ gates.  Table \ref{tab:resources} summarizes the resources required for the variants of Grover's algorithm.  
}

Numerically evolving (\ref{trottered}) and evaluating the success probability (\ref{fidelity}), we find there is a close similarity between the evolution of the continuous case version (see Fig. \ref{fig2} and \ref{fig3}), in terms of the timescale of the decay, as well as the revival times. For the parameters shown in Fig.~\ref{fig2}(d), the stroboscopic dynamics suppress some of the residual oscillatory behavior observed in the corresponding continuous-time curve in Fig.~\ref{fig2}(b). We consider this to be a side effect of the small register sizes making the discretization effects stronger.  For larger system sizes the continuous and discretized versions are more similar in evolution, as can be seen in Fig. \ref{fig3}.  

\textcolor{black}{We note that there is a frequency difference with a factor  $ \pi/2 $ between the continuous time evolution and discretized evolution curves in Fig. \ref{fig3} for the standard Grover case ($r=0$).  This can be attributed to the discrepancy between the continuous and discrete dynamics.  From Appendix \ref{app:standard}, the maximum fidelity is attained for the continuous model occurs at a time $ t = (\pi/2)\sqrt{N/M} $.  Meanwhile, in discrete case, the optimal number of Grover iterations is $ (\pi/4)\sqrt{N/M} $. Taking each Trotter timestep to be $ \delta  t = \pi $, this gives an additional factor of $ \pi/2 $. }

\section{Robustness against control errors}

\begin{figure}[t]
\includegraphics[width=\columnwidth]{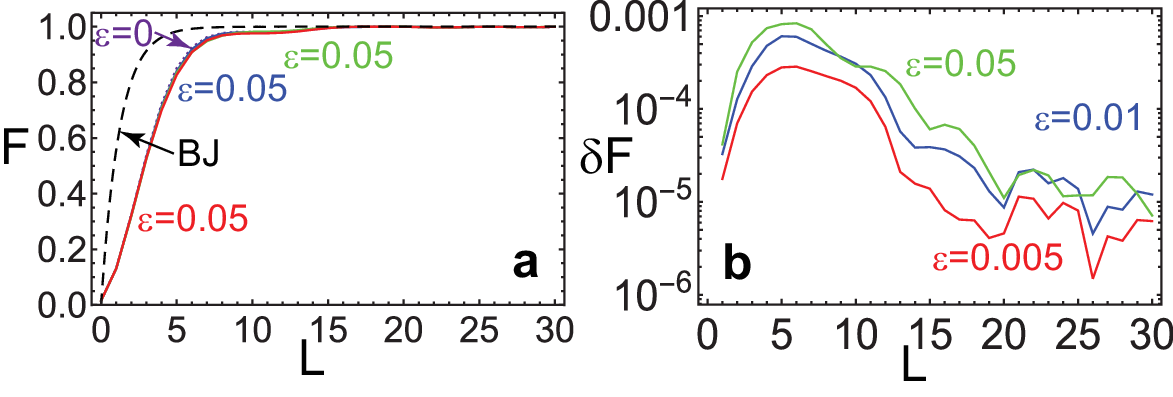}
\caption{
Effect of gate control errors on dissipative Grover's algorithm.  We 
perform (\ref{trottered}) where the time step $ \delta t $ in the gates (\ref{uplusdef}) and (\ref{usdef}) are randomly adjusted by a relative fraction $ \epsilon $.  (a) Three random instances with $ \epsilon = 0.05 $ (solid lines) as well as the ideal evolution $ \epsilon = 0 $ (dotted line). The evolution according to the BJ model is also shown (dashed line).  (b) The mean deviation $ \delta F = E[ | F(\epsilon) - F(\epsilon=0) | ] $ evaluated over 100 independent runs.  We use parameters $ M =1 $, $ n = 6 $, $ r = 3 $, $ \Delta =3 \sqrt{M(N-M)}/NR $ for all plots. 
\label{figerror}
}
\end{figure}

One of the advantages of our scheme is that it is robust against control errors due to the dissipative dynamics. Dissipation is a generic phenomenon which can happen in a variety of circumstances, and does not depend upon precisely tuned parameters. In Fig. \ref{figerror}, we show the effect of control errors on the gates applied in the sequence (\ref{trottered}).  We see in Fig. \ref{figerror}(a) even with 5\% control errors there is very little effect to the evolution.  In Fig. \ref{figerror}(b) we evaluate the mean deviation which quantifies the differences to the control error-free case.  We see that the most vulnerable part of the evolution is the decay itself, and once it has converged the effect of control errors is typically very small.  This is in contrast to fixed point methods (see Appendix \textcolor{black}{\ref{app:control}}) where a comparable evolution exhibits significant deviations to the final fidelity, due to the precisely optimized gates which must be applied.

\section{Conclusions}

We have introduced a variant of Grover's algorithm for performing an unstructured search that exhibits approximately exponential convergence toward the solution states over a broad pre-revival time window, rather than the sharply timed oscillations of the standard algorithm. The timescale of the exponential dynamics, for an optimally chosen $ \Delta, R $ is $ O( \sqrt{N/M} ) $,  matching the scaling of standard Grover's algorithm, up to reasonable prefactors.   We have formulated both the continuous and discrete time versions of the algorithm and shown that they work well with the expected exponential dynamics. One immediate application of our variant is that the precise number of Grover iterations does not need to be performed.   In conventional Grover's algorithm, one requires prior knowledge of the number of solutions $ M $ so that an optimal number of Grover iterations can be performed.  Here, as long as revival times are avoided, \textcolor{black}{approximately} exponential convergence towards the solution is obtained.  Since the revival times $ \tau $ only depend upon $ \Delta $, in practice, these times can be easily avoided. The scaling in the unknown $ M $ case is $ O( \sqrt{N} ) $; hence, either way, there is a  quantum speedup. This method is far less sensitive to control errors than existing methods, such as fixed point methods, which rely on a precise sequence of gates.  The trade-off is an increased number of qubits due to the ancilla reservoir and a slightly more complex iterate, which may nonetheless be offset in architectures where static Hamiltonians and reservoir engineering are more natural than long sequences of precisely calibrated phase rotations. We anticipate that this perspective may be useful for implementing robust search primitives on near-term hardware and for exploring connections between quantum algorithms and open-system dynamics.

% %=========================================================================

\section*{Acknowledgments}

This work is supported by the SMEC Scientific Research Innovation Project (2023ZKZD55); the Science and Technology Commission of Shanghai Municipality (22ZR1444600); the NYU Shanghai Boost Fund; the China Foreign Experts Program (G2021013002L); the NYU-ECNU Institute of Physics at NYU Shanghai; the NYU Shanghai Major-Grants Seed Fund; and Tamkeen under the NYU Abu Dhabi Research Institute grant CG008.

\section*{Data availability statement}
\textcolor{black}{All numerical calculations are fully reproducible using the equations provided. The code that support the findings of this study are openly available at the following URL: \url{https://github.com/J-Raghoonanan/Dissipative_Grover}}

% \section*{Author contributions}
% \textcolor{black}{S. C. and J. R. performed the calculations and analysis.  T. B. conceived the idea.  All authors wrote the paper.  }

\appendix

\section{Standard Grover's algorithm in continuous time}
\label{app:standard}

Here we discuss the solution of standard Grover's algorithm in continuous time given by (\ref{standardgroverham}).  We first transform the solution states into a Fourier basis defined as
\begin{align}
|\tilde{S}_p \rangle = \frac{1}{\sqrt{M}} \sum_{m=0}^{M-1} e^{i \frac{2 \pi}{M} p m } |S_m \rangle .
\end{align}
These states form an orthogonal set according to $ \langle \tilde{S}_p | \tilde{S}_{p'} \rangle = \delta_{p p'} $.  The Hamiltonian (\ref{standardgroverham}) can then be rewritten as
\begin{align}
H_\text{G} = |+^n \rangle\langle +^n| + \sum_{p=0}^{M-1} |\tilde{S}_p \rangle\langle\tilde{S}_p|  .
\label{fouriergroverham}
\end{align}
The state $ | +^n \rangle $ is a superposition of all states and hence involves some solution states $ |S_m \rangle $.  We may define an orthogonal basis by subtracting out the solution states from $ |+_n \rangle $
\begin{align}
|\perp^n \rangle &  := \sqrt{\frac{N}{N-M}} \left( | +^n \rangle- \frac{1}{\sqrt{N}} \sum_{m=0}^{M-1} | S_m \rangle \right) \label{perpdef} \\
& =\frac{1}{\sqrt{N-M}} \sum_{m=0}^{N-M-1} | \cancel{S}_m \rangle 
\end{align}
where $  | \cancel{S}_m \rangle  $ is the complement of the solution states (i.e. any state that is not a solution).  As such, it is orthogonal to the solution states
\begin{align}
\langle S_m |\perp^n \rangle  =  \langle \tilde{S}_p |\perp^n \rangle  = 0 ,
\end{align}
for $ m,p  \in [0, M-1] $. We may equivalently write (\ref{perpdef}) in terms of the Fourier basis states as
\begin{align}
| +^n \rangle = \sqrt{\frac{M}{N}} | \tilde{S}_0 \rangle + \sqrt{\frac{N-M}{N}} |\perp^n \rangle . \label{plusstatedef}
\end{align}
Substituting (\ref{plusstatedef}) into (\ref{fouriergroverham}), we obtain
%
% {\color{blue}
\begin{align}
H_\text{G} & = (1+ \frac{M}{N}) |\tilde{S}_0 \rangle \langle \tilde{S}_0 | + (1 -  \frac{M}{N})  | \perp^n \rangle \langle \perp^n | \nonumber \\
& + \sqrt{\frac{M(N-M)}{N^2}} ( | \tilde{S}_0 \rangle \langle \perp^n | +  | \perp^n \rangle \langle \tilde{S}_0  |)  \nonumber \\
& + \sum_{p=1}^{M-1} | \tilde{S}_p \rangle \langle \tilde{S}_p | .
\label{orthogonalizedhamgrover}
\end{align}
% }
%
The initial state in Grover's algorithm is $ | \psi_0 \rangle = | +^n \rangle $, which, from (\ref{plusstatedef}), we see can be written in terms of $ | \tilde{S}_0 \rangle $ and $ |\perp^n \rangle $.  Hence, the dynamics will be purely in terms of this two dimensional subspace and the last term in (\ref{orthogonalizedhamgrover}) plays no further role.  Diagonalizing the Hamiltonian (\ref{orthogonalizedhamgrover}), we obtain
\begin{align}
H_\text{G} & = \epsilon_+ | \epsilon_+  \rangle \langle \epsilon_+  | 
+ \epsilon_- | \epsilon_-  \rangle \langle \epsilon_-  | 
+ \sum_{p=1}^{M-1} | \tilde{S}_p \rangle \langle \tilde{S}_p |, 
\end{align}
where 
\begin{align}
| \epsilon_\pm  \rangle =  \frac{\sqrt{M(N-M)} |\tilde{S}_0 \rangle - (M \pm \sqrt{MN} )    | \perp^n \rangle}{\sqrt{2M(N \pm \sqrt{MN})}}
\end{align}
and
\begin{align}
\epsilon_\pm = 1 \pm \sqrt{\frac{M}{N} } .
\end{align}
For $ M\ll N $ we may approximate
\begin{align}
| \epsilon_\pm  \rangle & \approx   \frac{ |\tilde{S}_0 \rangle \mp     | \perp^n \rangle}{\sqrt{2}}  \label{approxenergystate} \\
|\perp^n \rangle  &  \approx  | +_n \rangle .
 \label{approxenergy} 
\end{align}
Then the time evolution of the initial state $ |\psi_0 \rangle =  | +^n \rangle $ can be approximated as
\begin{align}
|\psi(t) \rangle & = e^{-i \epsilon_+ t} \langle \epsilon_+  | \psi_0 \rangle   | \epsilon_+  \rangle  + e^{-i \epsilon_- t}\langle \epsilon_-  | \psi_0 \rangle   | \epsilon_-  \rangle \nonumber \\
& \approx - \frac{e^{-i \epsilon_+ t} }{\sqrt{2}} | \epsilon_+  \rangle  + 
 \frac{e^{-i \epsilon_- t} }{\sqrt{2}} | \epsilon_-  \rangle  \nonumber \\
 & = e^{-it} \left(  i\sin ( \sqrt{\frac{M}{N} }  t)  |\tilde{S}_0 \rangle    + \cos ( \sqrt{\frac{M}{N} }  t)  | \perp_n \rangle \right),
\end{align}
where we used (\ref{approxenergystate}) and (\ref{approxenergy}).  We see that at a time
\begin{align}
t_{\max} = \frac{\pi}{2} \sqrt{\frac{N}{M} }
\end{align}
the fidelity of the state in the solution space (\ref{fidelity}) reaches $ F = 1 $, which recovers the Grover scaling.

\section{The Bixon-Jortner model: Infinite reservoir solution}
\label{app:bjmodel}

\begin{figure}[t]
\includegraphics[width=\columnwidth]{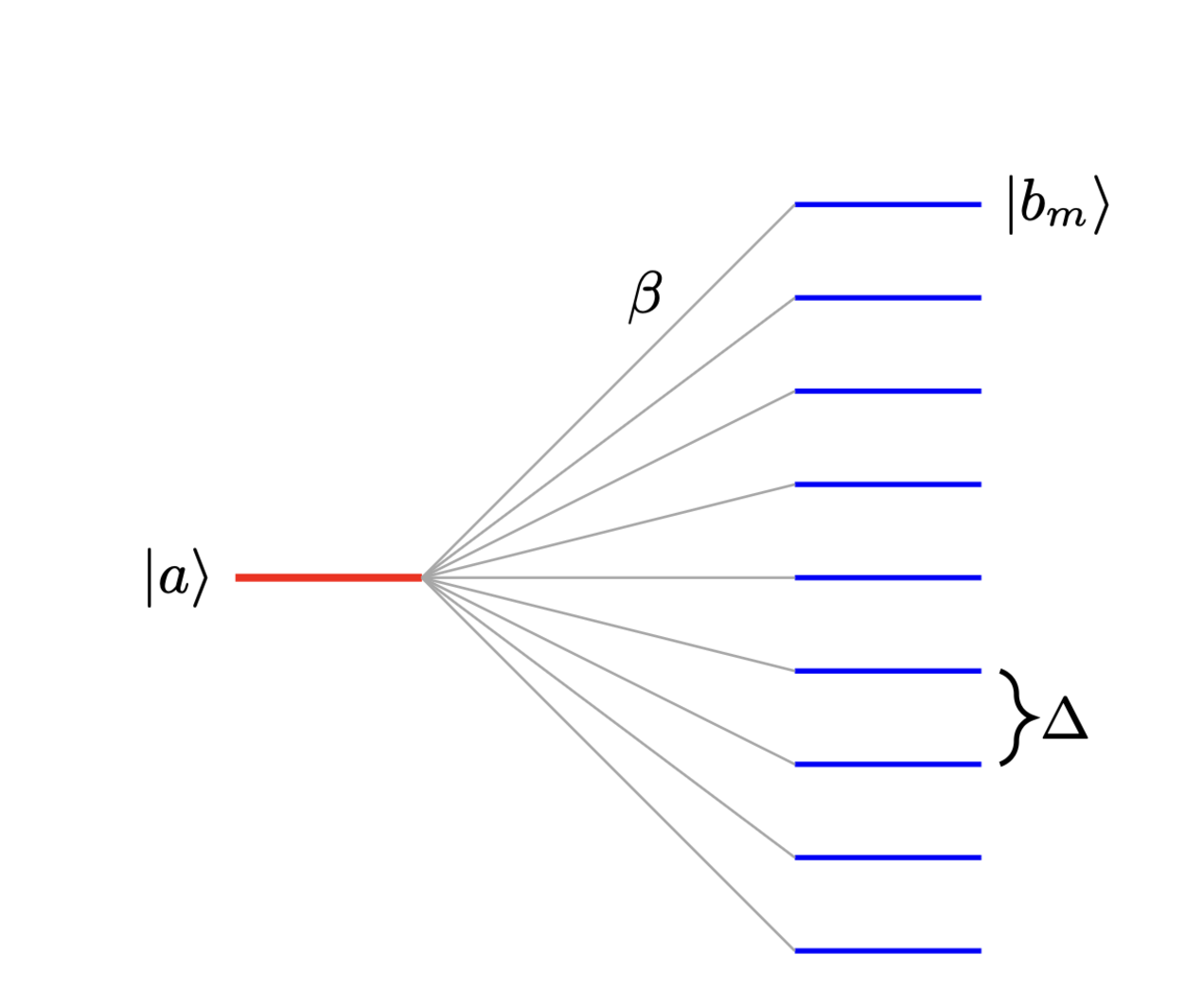}
    \caption{Energy levels of the Bixon-Jortner model. The state $\ket{a}$ is coupled to an infinite ladder of states  $\ket{b_m}$ via uniform coupling $\beta$.  The reservoir levels are regularly spaced by the energy $\Delta$.}
\label{fig:bj_model}
\end{figure}

The Bixon-Jortner (BJ) model (also called the row-column model) consists of a source state coupled to a reservoir and can be described by the following Hamiltonian \cite{bixon1968intramolecular}:
\begin{align}
H_{\text{BJ}} = \epsilon_a |a\rangle\langle a| + \sum_{m=-\infty}^\infty \left[ \epsilon_m |b_m \rangle\langle b_m| + \beta (|a \rangle\langle b_m| + |b_m \rangle\langle a|) \right]
\label{bjham}
\end{align}
where 
\begin{align}
\epsilon_m = \epsilon_a+ m\Delta
\end{align}
for $ m \in \mathbb{Z}$.  Here the state $ | a \rangle $ is the source state and the states $ | b_m \rangle $ are the reservoir states (see Fig. \ref{fig:bj_model}).

% \begin{figure}[t]
% \centering
% \begin{tikzpicture}
% % Parameters
% \def\nlevels{9}
% \def\spacing{0.5}
% % Draw central state |a,0>
% \draw[very thick, red] (-1,2.5) -- (0,2.5);
% \node[left] at (-1,2.5) {$\ket{a}$};
% % Draw ladder states |g, nΔ⟩
% \foreach \i in {1,...,\nlevels} {
%     \draw[thick, blue] (2,\i*\spacing) -- (3,\i*\spacing); % horizontal lines
%     \draw[gray!70] (0,2.5) -- (2,\i*\spacing); % connections
% }
% % Labels
% \node[right] at (3,\nlevels*\spacing) {$\ket{b_m}$};
% \node at (1.2, 4) {$\beta$};
% % Bracket for spacing Δ
% \draw[decorate, decoration={brace, mirror, amplitude=5pt}, thick] (3.1,3*\spacing) -- (3.1,4*\spacing);
% \node[right] at (3.2,3.5*\spacing) {$\Delta$};
% \end{tikzpicture}
% \caption{Energy levels of the Bixon-Jortner model. The state $\ket{a}$ is coupled to an infinite ladder of states  $\ket{b_m}$ via uniform coupling $\beta$.  The reservoir levels are regularly spaced by the energy $\Delta$.}
% \label{fig:bj_model}
% \end{figure}

The BJ model is typically initialized in the state $ | \psi_0 \rangle = |a \rangle $ and describes the 
transition to the reservoir states. In the case of an infinite number of states $\ket{b_m}$, the behavior of this system is analogous to the dynamics of the continuum Fano--Anderson model in the limit where the coupling parameter is kept constant and the level spacing $\Delta \rightarrow 0$, describing an excited discrete level coupled to a continuum of modes~\cite{Fano1961,Anderson1958}. For short time evolutions, it can be approximated using Fermi's ``golden rule'': the probability amplitude of the initial state is characterized by an exponential decay $e^{-\gamma t}$ in the time period $(0, \tau)$, where $\tau$ characterizes a revival time. 

The Schr\"odinger equation yields the following system of differential equations 
\begin{align}
\frac{da}{dt} & = -i\epsilon_a a(t) -i\beta \sum_{m=-\infty}^\infty b_m(t) \nonumber \\
\frac{db_m}{dt} & = -i\epsilon_m b_m(t) - i\beta a(t) .
\label{systemeqbj}
\end{align}
The dynamics can be solved exactly in the infinite reservoir case where the source-state amplitude is \cite{hansen2023decay}  

\begin{align}
  a(t) = &  e^{-i\epsilon_a t} \Big[ e^{-\gamma t/2} - \sum_{k=1}^\infty \frac{\gamma (t- k \tau)}{k} e^{-\gamma(t-k\tau)/2} \nonumber \\
  & \times \mathcal{L}_{k-1}^{ (1)}  (\gamma(t-k\tau))\Theta(t-k\tau)   \Big] 
  \label{exactsolbj}
\end{align}

where $ \mathcal{L}_{n}^{(\alpha )}  (x) $ is the generalized Laguerre polynomial and the decay rate is 
\begin{align}
    \gamma = 2\pi\frac{\beta^2}{\Delta} ,
    \label{gammadef}
\end{align}
and the revival time is 
\begin{align}
    \tau = \frac{2\pi}{\Delta} .
    \label{taudef}
\end{align}
We note that  $\epsilon_a$ only contributes a global phase so does not affect the dynamics.  Specifically, in the time range $ 0 \le t < \tau $ the amplitude obeys an exponential evolution
\begin{align}
  a(t) = e^{-i\epsilon_a t} e^{-\gamma t/2} .
\end{align}
For longer times, there are revivals due to the Laguerre polynomial that ``kick in'' every time a multiple of $\tau$ has elapsed.

\section{The Bixon-Jortner model: Finite reservoir approximation}
\label{app:bjmodelfinite}

\begin{figure}[t]
\includegraphics[width=\columnwidth]{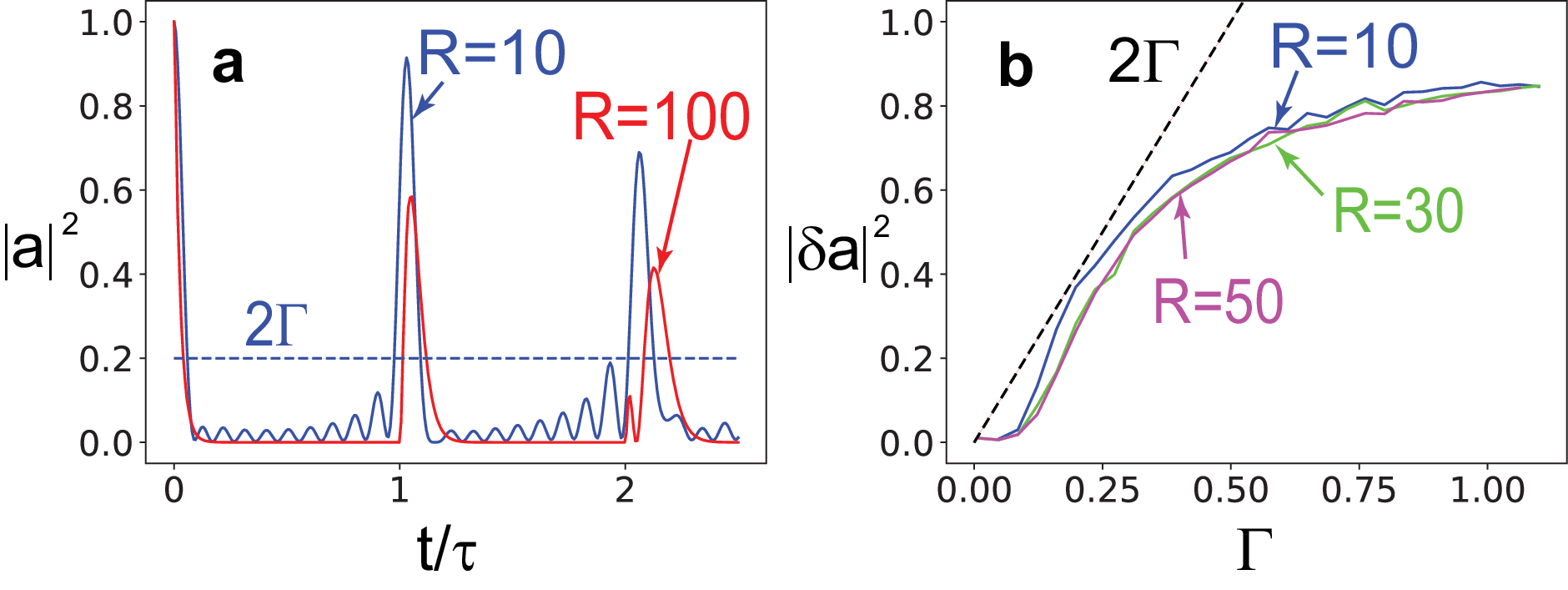}
    \caption{(a) Evolution of $|a(t)|^2$ under the finite reservoir BJ model (\ref{finitebjham}) for different values of $R$. We plot two curves with the parameters  $R = 10$ (corresponding to $\Gamma = \frac{\beta^2}{\Delta^2 R} = 0.1$) and $R=100$ 
    (corresponding to $\Gamma = 0.01 $).  The bound $ 2 \Gamma $ is shown for the parameters with $ R =10 $.  
 (b) Amplitude of the residual oscillations $ \delta a $ versus  $ \Gamma $.  The solid lines represents the maximum amplitude computed between the initial decay and the revival time in simulations for $R = 10, 30, 50$. 
 The dashed line corresponds to the bound $ |\delta a |^2  = 2\Gamma $ given  in (\ref{bounddela}).  We use  $\Delta = 1$, $\beta = 1$ for all simulations.  
 }.
    \label{fig:BJevolution}%
\end{figure}

In dissipative Grover's algorithm, the size of the reservoir is always finite as it consists of ancilla qubits.  We define the finite reservoir Bixon-Jortner model as
\begin{align}
& H_{\text{FBJ}} =  \epsilon_a |a \rangle\langle a| \nonumber \\
& + \sum_{m=-R/2}^{R/2} \left[ \epsilon_m |b_m \rangle\langle b_m| + \beta (|a \rangle\langle b_m| + |b_m \rangle\langle a|) \right]
\label{finitebjham}
\end{align}
In the finite-reservoir case, there is no analytical solution to our knowledge.  The main effect that arises from the truncation of the reservoir is residual oscillations as shown in Fig. \ref{fig:BJevolution}. In this section we estimate the magnitude of the fluctuations due to the finite reservoir.  

First, by making the substitution 
\begin{align}
   a(t) & =  e^{-i \epsilon_a t} \tilde{a}(t) \nonumber \\
   b_m(t) & =  e^{-i \epsilon_a t} \tilde{b}_m(t)  
\end{align}
in (\ref{finitebjham}), we may eliminate the energy term on $ |a \rangle $, and have the equivalent equations
\begin{align}
\frac{d \tilde{a} }{dt} & =  -i\beta \sum_{m}^{}  \tilde{b}_m(t) \label{akeqn}
\\
\frac{d \tilde{b}_m}{dt} & = -i m \Delta   \tilde{b}_m(t) - i\beta \tilde{a} (t) . \label{bkeqn}
\end{align}
where the sum is over reservoir states. From the exact solution (\ref{exactsolbj}) in the time range $ 0< t < \tau $, we know that $ \tilde{a} (t) = e^{-\gamma t/2} $ in the infinite reservoir limit.  Substituting this into (\ref{bkeqn}) we obtain
\begin{align}
\frac{d \tilde{b}_m}{dt} & = -i m \Delta   \tilde{b}_m(t) - i\beta e^{-\gamma t/2} 
\end{align}
which can be integrated to give
\begin{align}
\tilde{b}_m(t)  = \frac{2i \beta}{\gamma - 2i m \Delta } ( e^{-\gamma t/2} - e^{-im \Delta t})  ,
\end{align}
where we have chosen the integration constant such that $\tilde{b}_m(t=0)  = 0  $.  Substituting this into (\ref{akeqn}), we have
\begin{align}
    \frac{d \tilde{a} }{dt} & =  2 \beta^2 \sum_{m=-R/2}^{R/2} \frac{e^{-\gamma t/2} - e^{-i m \Delta t} }{\gamma - 2i m \Delta }   .
    \label{dadteqn}
\end{align}
We next approximate the sum in the above equation by turning sums into integrals. The first term in (\ref{dadteqn}) is 
\begin{align}
I_1 & = \sum_{m=-R/2}^{R/2} \frac{1}{\gamma - 2i m \Delta } \approx \int_{-R/2}^{R/2} d m \frac{1}{\gamma - 2i m \Delta }  \nonumber \\ 
& = \frac{1}{\Delta} \arctan (\frac{\Delta R}{\gamma} ) .
\end{align}
The second term in (\ref{dadteqn}) is meanwhile
\begin{align}
I_2  & =  \sum_{m=-R/2}^{R/2} \frac{e^{-i m \Delta t} }{\gamma - 2i m \Delta } 
  \approx  \int_{-R/2}^{R/2} d m  \frac{e^{-i m \Delta t} }{\gamma - 2i m \Delta }  \nonumber \\ 
 & = \frac{i e^{-\gamma t/2}}{2 \Delta} \left[ \text{Ei} ( \frac{(\gamma - i \Delta R)t }{2}) - \text{Ei} ( \frac{(\gamma + i \Delta R)t }{2})  \right] ,
\end{align}
where $ \text{Ei} (z) $ is the exponential integral function.  Using these definitions (\ref{dadteqn}) becomes
\begin{align}
    \frac{d \tilde{a} }{dt} & =   2 \beta^2 ( e^{-\gamma t/2} I_1 - I_2) .
    \label{dadteqn2}
\end{align}

Here, let us check that (\ref{dadteqn2}) reproduces the correct result for large  $R $.  In the limit $ R \rightarrow \infty $, we have according to the expressions above
\begin{align}
\lim_{R \rightarrow \infty}   I_1 & = \frac{\pi}{2 \Delta}  \\
\lim_{R \rightarrow \infty}   I_2 & =\frac{\pi}{\Delta}  e^{-\gamma t/2} ,
\end{align}
where we used the fact that $ \lim_{x \rightarrow \infty} \text{Ei} (ix) = i \pi \text{sgn} (x) $.  Substituting these into (\ref{dadteqn2}), we have
\begin{align}
  \frac{d \tilde{a} }{dt} & = - \frac{\pi \beta^2 }{\Delta}  e^{-\gamma t/2} ,
\end{align}
which can be readily integrated with $ t $ to confirm that $ \tilde{a}(t)  = e^{-\gamma t/2}  $ using (\ref{gammadef}).  This gives a sanity check that our methods are consistent. 

Now let us obtain the first order deviations from the large $ R $ limit.  For the first quantity we have
\begin{align}
I_1 & = \frac{1}{\Delta} ( \frac{\pi}{2} +  \arctan (\frac{\Delta R}{\gamma} ) - \frac{\pi}{2} )  \nonumber \\
& = \frac{1}{\Delta} ( \frac{\pi}{2} -  \arctan (\frac{\gamma}{\Delta R} ) )\nonumber \\ 
& \approx \frac{1}{\Delta}( \frac{\pi}{2} - \frac{\gamma}{\Delta R} ) .
\label{approxi1}
\end{align}
For the second quantity we may approximate for large $ R $ 
\begin{align}
I_2  & \approx \frac{i e^{-\gamma t/2}}{2 \Delta} \left[ \text{Ei} ( \frac{- i \Delta R t }{2}) - \text{Ei} ( \frac{i \Delta Rt }{2})  \right] \nonumber \\
& =  \frac{\pi  e^{-\gamma t/2}}{ \Delta}  -  \frac{2 e^{-\gamma t/2}}{  \Delta^2 R t} 
\cos (\frac{\Delta R t}{2}) ,
\label{approxi2}
\end{align}
where we used the approximation for the exponential integral with a purely imaginary argument
\begin{align}
\text{Ei} (iy) \approx i \pi \text{sgn} (y) + \frac{e^{iy}}{iy} .
\end{align}
Substituting the approximate expressions (\ref{approxi1}) and (\ref{approxi2}) into (\ref{dadteqn2}), we have
\begin{align}
 \frac{d \tilde{a} }{dt} & =  - \frac{\pi \beta^2 }{\Delta}  e^{-\gamma t/2} - \frac{2 \beta^2 \gamma}{\Delta^2 R } e^{-\gamma t/2} + 
 \frac{4 \beta^2  e^{-\gamma t/2}}{  \Delta^2 R t} 
\cos (\frac{\Delta R t}{2}) .
\end{align}
Integrating with $ t $, we have
\begin{align}
\tilde{a}(t) & =  e^{-\gamma t/2} - \frac{4 \beta^2 }{\Delta^2 R }e^{-\gamma t/2} \nonumber \\
& +  \frac{2 \beta^2 }{\Delta^2 R} \left[ \text{Ei} ( - \frac{(\gamma - i R \Delta)t }{2} ) + \text{Ei} ( - \frac{(\gamma + i R \Delta)t }{2} )  \right]  .
\end{align}
We see that the correction terms to the dominant exponential term are of order 
\begin{align}
    \delta \tilde{a}(t) \sim \frac{\beta^2}{\Delta^2 R} 
    \label{residualamp}
\end{align}
as claimed in the main text.  

As shown in Fig. \ref{fig:BJevolution}, the amplitude of the oscillations is indeed mainly given by the ratio $\Gamma = \beta^2/\Delta^2R$. Using the maximum amplitude computed in the numerical simulations, we can see that 
\begin{align}
    | \delta \tilde{a}_{\max} |^2  \le  2\Gamma = \frac{2 \beta^2}{\Delta^2R } 
    \label{bounddela}
\end{align}
gives a bound for the amplitude of the oscillations.

\section{Mapping to dissipative Grover Hamiltonian}
\label{app:mapping}

Comparing the Hamiltonian (\ref{bjlikeham}) to (\ref{finitebjham}) and associating $ |a \rangle \leftrightarrow | \perp^n, +^r \rangle $, $ |b_m \rangle \leftrightarrow |\tilde{S}_0,k \rangle $, $ m \leftrightarrow k-R/2 $ we have
\begin{align}
\epsilon_a & = 1 - \frac{M}{N} \nonumber \\
\epsilon_{k-R/2} & = E_k  {\color{black} +  \frac{M}{NR}} \nonumber \\
\beta & = \frac{1}{N}\sqrt{\frac{M(N-M)}{R}} .
\end{align}
The reservoir energy spacing $ \Delta $ is the same for both models.  

Neglecting the last line in (\ref{bjlikeham}) and in the limit of $ R \rightarrow \infty$, the evolution is given by (\ref{exactsolbj}), with parameters (\ref{gammadef}) and (\ref{taudef}). \textcolor{black}{In the ideal BJ model initialized purely in the source state $\ket a$, the reservoir population is}
\begin{align}
% \textcolor{blue}{
F_{\rm res}(t) = \sum_m |b_m(t)|^2 = 1-|a(t)|^2 .
% }
\end{align}
%
% \textcolor{blue}{
In the dissipative Grover case, the state does not start in the state corresponding to $ |a \rangle $.  The initial state is
\begin{align}
|+^n, +^r \rangle & = \sqrt{\frac{M}{N}} | \tilde{S}_0, +^r \rangle + \sqrt{\frac{N-M}{N}} | \perp^n, +^r \rangle \\
& \leftrightarrow \sqrt{\frac{M}{NR}} \sum_{k=0}^{R-1} | b_{k-R/2} \rangle + \sqrt{\frac{N-M}{N}} | a \rangle . 
\end{align}
Hence the source-state weight is $(N-M)/N$, while the total reservoir weight is $M/N$. Since we primarily consider $ M \ll N $, we make the approximation that only the initial source state weight undergoes the BJ dynamics, and the initial reservoir population stays in the reservoir.  The physical success probability consistent with (\ref{fidelity}) is
% }

%
% \textcolor{blue}{
\begin{align}
F_{\text{BJ}} (t) & =  \frac{M}{N} +  \frac{N-M}{N} \left(1-|a(t)|^2\right) \nonumber\\
&=
1-\frac{N-M}{N}|a(t)|^2 .
\label{fidbj}
\end{align}
%
% }

For finite $ R $, the size of the residual oscillations is given by (\ref{residualamp})
\begin{align}
\Gamma = \frac{\beta^2}{\Delta^2 R} =  \frac{M(N-M)}{ ( R N \Delta )^2 } .
\end{align}

% {\color{blue}
\section{Decay time scaling using the Bixon-Jorter model for unknown-$ M $}
\label{app:unknown}

% Equivalently, one can choose a measurement time based on the slowest sparse case, $M=1$.  For example, taking
% \begin{equation}
%     t_{\mathrm{meas}} = \alpha \sqrt{\frac{RN}{C}},
% \end{equation}
% with $\alpha=O(1)$, gives, for $M=1$,
% \begin{equation}
%     \gamma t_{\mathrm{meas}}
%     =
%     \alpha\frac{N-1}{N}.
% \end{equation}
% so $t_{\mathrm{meas}}$ gives a failure probability approximately $e^{-\alpha}$ for $M=1$, and a smaller failure probability for larger $M$ in the sparse regime.

In the BJ approximation, the physical failure probability is
\begin{equation}
\label{BJ_failure}
    1-F_{\rm BJ}(t)
    =
    \frac{N-M}{N}e^{-\gamma t},
\end{equation}
where we used (\ref{fidbj}) and (\ref{abj}) before the first revival time.  Finding the time to reach $ F_{\text{BJ}} = 0.99 $ for $M=1$, we obtain
\begin{align}
    t_{F=0.99}^{\rm BJ}
    &= -\frac{1}{\gamma} \log\!\left[\frac{1-0.99}{(N-1)/N}\right] \nonumber\\
    &\approx
    \log(100)\sqrt{\frac{RN}{C}}, 
\end{align}
where we used (\ref{gammabj}) and (\ref{deltaunknownchoice}) with $ M =1 $ and discarded $O(1/N)$ corrections.

% }

\section{Sensitivity to control errors}
\label{app:control}

To simulate the effect of imperfect gates, we add a control error to the phases in the Grover iterate operators as
\begin{align}
U_+ & = e^{-i |+^n,+^r \rangle\langle +^n,+^r| (1+ \epsilon \xi ) \delta t}  \\
U_{S} & = e^{-i \sum_m \sum_k E_k |S_m,k \rangle\langle S_m,k| (1+ \epsilon \xi' )  \delta t}  ,  \\
\end{align}
where $ \xi, \xi' $ are uniformly distributed random variables in the range $ [-1,1] $.  We then perform the evolution (\ref{trottered}), where the random variables $ \xi, \xi' $ are chosen differently each time the gate is implemented.  

Figure \ref{figerror}(a) shows the effect of control errors on the performance of our algorithm.  As expected the control errors create a deviation of the fidelity from the ideal case.  We estimate the effect of the control errors on the fidelity by finding the mean deviation with respect to the error-free fidelity:
\begin{align}
\delta F = E[ | F(\epsilon) - F(\epsilon=0) | ] ,
\label{meandev}
\end{align}
where $ E[ \cdot ] $ denotes the expectation value over independent runs of the algorithm.  Results are shown in Fig. \ref{figerror}(b).  We find that for small errors the the mean deviation of the fidelity stays approximately constant with the number of gate operations.

We contrast this with the fixed-point algorithm approach of Ref. \cite{yoder2014fixed}.  The gate sequence that is applied in this case is
\begin{align}
|\psi_{\text{FP}} (\ell) \rangle = \left[ \prod_{j=1}^\ell G(\alpha_j (1+ \epsilon \xi  ) , \beta_j (1+ \epsilon \xi' ) ) \right]  | \psi_0 \rangle ,
\label{fpsequence}
\end{align}
where the Grover iterate is defined as
\begin{align}
 G(\alpha, \beta  ) = - S_s (\alpha) S_t (\beta) 
\end{align}
and
\begin{align}
S_s (\alpha) & = I - (1- e^{-i\alpha}) | + \rangle \langle + |^{\otimes n} \\
S_t (\beta) & =  I - (1- e^{i \beta}) \sum_{m=0}^{M-1} | S_m \rangle \langle S_m  | .
\end{align}
The optimal angles are given by
\begin{align}
\alpha_j & = - \beta_{\ell-j+1} \nonumber \\
& = 2 \cot^{-1} \left[  \tan (\frac{2 \pi j}{2\ell+1}) \sqrt{ 1- \frac{1}{T_{1/(2\ell+1)}^2 (1/\delta)} } \right] ,
\end{align}
where $ T_m (x) = \cos [ m \cos^{-1} (x) ] $ is the $m$th Chebyshev polynomial of the first kind.  Here $ \delta $ is an accuracy parameter which allows the fidelity $ F \ge 1- \delta^2 $ at the end of the evolution, in the noise-free case.  The number of iterations $ l $ must be chosen large enough to guarantee this, which is performed by setting
\begin{align}
   2\ell+1 \ge  \frac{\log (2/\delta)}{\sqrt{M/N}} .
   \label{ldeltarelation}
\end{align}

\begin{figure}[t]
\includegraphics[width=\columnwidth]{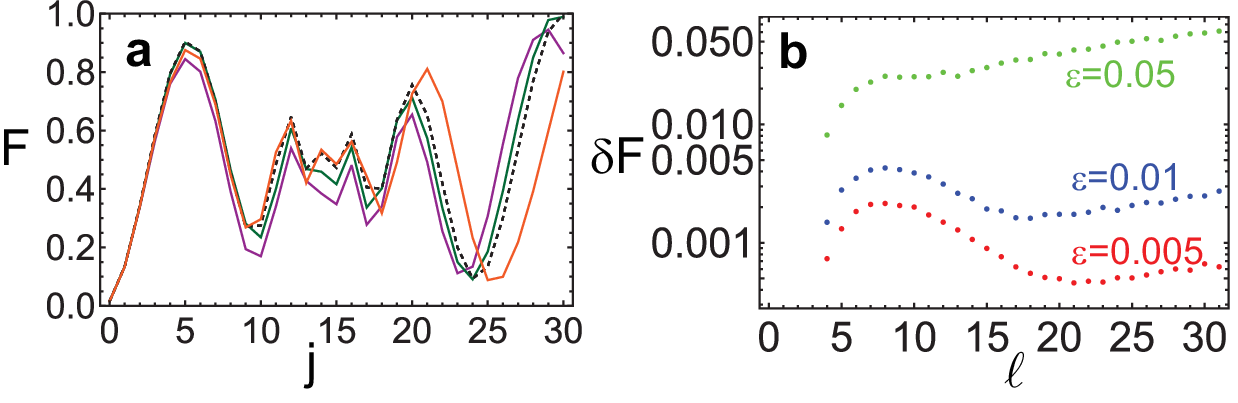}
\caption{
Effect of gate control errors on the fixed point algorithm of Ref. \cite{yoder2014fixed}.  (a) Fidelity (\ref{fidelity}) under the evolution (\ref{fpsequence}) for $ M = 1 $, $ N = 2^6 $.  Dashed line shows the ideal evolution ($ \epsilon = 0 $), solid lines show three trajectories with $ \epsilon = 0.05 $.  (b) The mean deviation (\ref{meandev}) as a function of the gate sequence length $ \ell $, where $ \delta $ is chosen according to (\ref{ldeltarelation}), by setting the inequality as an equality.  The control errors $ \epsilon $ in the sequence (\ref{fpsequence}) are as marked.  In (\ref{meandev}),  we take average over 1000 independent runs to estimate the mean deviation.  
\label{figerrorfp}
}
\end{figure}

In our calculations we fix $ M, N $, and choose $ \delta $ according to (\ref{ldeltarelation}) by setting the inequality to an equality.  In (\ref{fpsequence}) the variables $ \xi, \xi' \in [-1,1] $ are chosen randomly for each $ j$. 
Figure \ref{figerrorfp}(a) shows the typical evolution throughout the gate sequence.  We see that there is a much larger variation of the fidelity in comparison to Fig. \ref{figerror}(a).  This can be quantified in Fig. \ref{figerrorfp}(b) by evaluating the mean deviation (\ref{meandev}) at the end of the evolution, for various $ \ell $.

% How to do the references:
\bibliographystyle{apsrev}
\bibliography{NIIReferences}

@article{grover2005fixed,
  title={Fixed-point quantum search},
  author={Grover, Lov K},
  journal={Physical Review Letters},
  volume={95},
  number={15},
  pages={150501},
  year={2005},
  publisher={APS}
}

@article{martyn2021grand,
  title={Grand unification of quantum algorithms},
  author={Martyn, John M and Rossi, Zane M and Tan, Andrew K and Chuang, Isaac L},
  journal={PRX quantum},
  volume={2},
  number={4},
  pages={040203},
  year={2021},
  publisher={APS}
}

@article{hansen2023decay,
  title={Decay and revival dynamics of a quantum state embedded in a regularly spaced band of states},
  author={Hansen, Jan Petter and Tywoniuk, Konrad},
  journal={Physical Review A},
  volume={108},
  number={5},
  pages={053707},
  year={2023},
  publisher={APS}
}

@inproceedings{grover1996fast,
  title={A fast quantum mechanical algorithm for database search},
  author={Grover, Lov K},
  booktitle={Proceedings of the twenty-eighth annual ACM symposium on Theory of computing},
  pages={212--219},
  year={1996}
}

@article{bennett1997strengths,
  title={Strengths and weaknesses of quantum computing},
  author={Bennett, Charles H and Bernstein, Ethan and Brassard, Gilles and Vazirani, Umesh},
  journal={SIAM journal on Computing},
  volume={26},
  number={5},
  pages={1510--1523},
  year={1997},
  publisher={SIAM}
}

@book{nielsen2010quantum,
  title={Quantum computation and quantum information},
  author={Nielsen, Michael A and Chuang, Isaac L},
  year={2010},
  publisher={Cambridge university press}
}

@article{Durr96aquantum,
    author  = "Christoph D{\"u}rr and Peter H{\o}yer",
    title   = "A Quantum Algorithm for Finding the Minimum",
    year    = "1996",
      archivePrefix  = "arXiv",
      primaryClass   = "quant-ph",
    eprint   = "quantph/9607014"
}

@article{Magniez07,
author = {Frédéric  Magniez and Miklos  Santha and Mario  Szegedy},
title = {Quantum Algorithms for the Triangle Problem},
journal = {SIAM Journal on Computing},
volume = {37},
number = {2},
pages = {413-424},
year = {2007}
}

@article{Hsu03,
    author  = "Li-Yi Hsu",
    title   = "Quantum secret-sharing protocol based on Grover’s algorithm",
    year    = "2003",
    journal = "Phys. Rev. A",
    volume  = "68",
    number  = " ",
    pages   = "022306"
}

@INPROCEEDINGS{Brassard98,
    author = {Gilles Brassard and Peter H{\o}yer and Alain Tapp},
    title = {Quantum Counting},
    booktitle={Proceedings of 25th ICALP, Vol. 1443 of Lecture},
    year = {1998},
    pages = {820--831},
    publisher = {Springer}
}

@article{Chen05,
    author  = "Chen-Bin Zhang",
    title   = "Control of non-controllable quantum systems: a quantum control algorithm based on Grover iteration",
    year    = "2005",
    journal = "Journal of Optics B",
    volume  = "7",
    number  = " ",
    pages   = ""
}

@Article{Aimeur2013,
author="A{\"i}meur, Esma
and Brassard, Gilles
and Gambs, S{\'e}bastien",
title="Quantum speed-up for unsupervised learning",
journal="Machine Learning",
year="2013",
volume="90",
number="2",
pages="261--287"
}

@INPROCEEDINGS{Furer2008,
    author = {F{\"u}rer, Martin},
    title = {Quantum Counting},
    booktitle={LATIN 2008: Theoretical Informatics: 8th Latin American Symposium, B{\'u}zios, Brazil, April 7-11, 2008. Proceedings},
    year = {2008},
    pages = {784--792},
    publisher = {Berlin, Heidelberg}
}

@article{Bennett1997,
 author = {Bennett, Charles H. and Bernstein, Ethan and Brassard, Gilles and Vazirani, Umesh},
 title = {Strengths and Weaknesses of Quantum Computing},
 journal = {SIAM J. Comput.},
 issue_date = {Oct. 1997},
 volume = {26},
 number = {5},
 month = oct,
 year = {1997},
 pages = {1510--1523},
 numpages = {14},
 publisher = {Society for Industrial and Applied Mathematics},
 address = {Philadelphia, PA, USA},
}

@article{Brassard1997,
      author         = "Brassard, Gilles and H{\o}yer, Peter and Tapp, Alain",
      title          = "{Quantum algorithm for the collision problem}",
      year           = "1997",
      eprint         = "quant-ph/9705002",
      archivePrefix  = "arXiv",
      primaryClass   = "quant-ph",
      SLACcitation   = "%%CITATION = QUANT-PH/9705002;%%"
}

@article{Grover1997,
      author         = "Grover, Lov K.",
      title          = "{Quantum telecomputation}",
      year           = "1997",
      eprint         = "quant-ph/9704012",
      archivePrefix  = "arXiv",
      primaryClass   = "quant-ph",
      SLACcitation   = "%%CITATION = QUANT-PH/9704012;%%"
}

@Article{Hao2010,
author="Hao, Liang
and Li, JunLin
and Long, GuiLu",
title="Eavesdropping in a quantum secret sharing protocol based on Grover algorithm and its solution",
journal="Science China Physics, Mechanics and Astronomy",
year="2010",
volume="53",
number="3",
pages="491--495"
}

@INPROCEEDINGS{5954250,
author={Y. Wang and M. Perkowski},
booktitle={2011 41st IEEE International Symposium on Multiple-Valued Logic},
title={Improved Complexity of Quantum Oracles for Ternary Grover Algorithm for Graph Coloring},
year={2011},
pages={294-301},
month={May},}

@ARTICLE{2005quant.ph.4012A,
   author = {{Ambainis}, A.},
    title = "{Quantum search algorithms}",
   eprint = {quant-ph/0504012},
     year = 2005,
      archivePrefix  = "arXiv",
      primaryClass   = "quant-ph",
}

@article{yoder2014fixed,
  title={Fixed-point quantum search with an optimal number of queries},
  author={Yoder, Theodore J and Low, Guang Hao and Chuang, Isaac L},
  journal={Physical review letters},
  volume={113},
  number={21},
  pages={210501},
  year={2014},
  publisher={APS}
}

@article{byrnes2018generalized,
  title={Generalized Grover’s algorithm for multiple phase inversion states},
  author={Byrnes, Tim and Forster, Gary and Tessler, Louis},
  journal={Physical review letters},
  volume={120},
  number={6},
  pages={060501},
  year={2018},
  publisher={APS}
}

@article{bixon1968intramolecular,
  title={Intramolecular radiationless transitions},
  author={Bixon, Mordechai and Jortner, Joshua},
  journal={The Journal of chemical physics},
  volume={48},
  number={2},
  pages={715--726},
  year={1968},
  publisher={American Institute of Physics}
}

@article{liao2024quadratic,
  title={Quadratic quantum speedup for perceptron training},
  author={Liao, Pengcheng and Sanders, Barry C and Byrnes, Tim},
  journal={Physical Review A},
  volume={110},
  number={6},
  pages={062412},
  year={2024},
  publisher={APS}
}

@article{tessler2017bitcoin,
  title={Bitcoin and quantum computing},
  author={Tessler, Louis and Byrnes, Tim},
  journal={arXiv preprint arXiv:1711.04235},
  year={2017}
}

@article{grover2006quantum,
  title={Quantum algorithms with fixed points: The case of database search},
  author={Grover, Lov K and Patel, Apoorva and Tulsi, Tathagat},
  journal={arXiv preprint quant-ph/0603132},
  year={2006}
}

@article{Fano1961,
  title = {Effects of Configuration Interaction on Intensities and Phase Shifts},
  author = {Fano, U.},
  journal = {Phys. Rev.},
  volume = {124},
  issue = {6},
  pages = {1866--1878},
  numpages = {0},
  year = {1961},
  month = {Dec},
  publisher = {American Physical Society}
}

@article{Anderson1958,
  title = {Absence of Diffusion in Certain Random Lattices},
  author = {Anderson, P. W.},
  journal = {Phys. Rev.},
  volume = {109},
  issue = {5},
  pages = {1492--1505},
  numpages = {0},
  year = {1958},
  month = {Mar},
  publisher = {American Physical Society}
}

@article{Campbell2019CSP,
   title={Applying quantum algorithms to constraint satisfaction problems},
   volume={3},
   ISSN={2521-327X},
   journal={Quantum},
   publisher={Verein zur Forderung des Open Access Publizierens in den Quantenwissenschaften},
   author={Campbell, Earl and Khurana, Ankur and Montanaro, Ashley},
   year={2019},
   month=jul, pages={167} }

@article{Gao2022GWGrover,
  title = {Quantum algorithm for gravitational-wave matched filtering},
  author = {Gao, Sijia and Hayes, Fergus and Croke, Sarah and Messenger, Chris and Veitch, John},
  journal = {Phys. Rev. Res.},
  volume = {4},
  issue = {2},
  pages = {023006},
  numpages = {24},
  year = {2022},
  month = {Apr},
  publisher = {American Physical Society}
}

@article{Morimoto2024QuADS,
  title = {Continuous optimization by quantum adaptive distribution search},
  author = {Morimoto, Kohei and Takase, Yusuke and Mitarai, Kosuke and Fujii, Keisuke},
  journal = {Phys. Rev. Res.},
  volume = {6},
  issue = {2},
  pages = {023191},
  numpages = {11},
  year = {2024},
  month = {May},
  publisher = {American Physical Society}
}

@conference{Sarah2024PracticalGroverAES,
  title = {On the practical cost of Grover for AES key recovery},
  author = {Sarah D. and Peter C.},
    year = 2024,
    month = {March},
    booktitle = {NIST Fifth PQC Standardisation Conference},
    organization = {UK National Cyber Security Centre}
}

@article{Long_2001,
   title={Grover algorithm with zero theoretical failure rate},
   volume={64},
   number={2},
   journal={Physical Review A},
   publisher={American Physical Society (APS)},
   author={Long, G. L.},
   year={2001},
   month=jul }

@article{Morales_2018,
   title={Variational learning of Grover’s quantum search algorithm},
   volume={98},
   ISSN={2469-9934},
   number={6},
   journal={Physical Review A},
   publisher={American Physical Society (APS)},
   author={Morales, Mauro E. S. and Tlyachev, Timur and Biamonte, Jacob},
   year={2018},
   month=dec }

@article{Zhang2024,
  title = {Grover-QAOA for 3-SAT: quadratic speedup,  fair-sampling,  and parameter clustering},
  volume = {10},
  number = {1},
  journal = {Quantum Science and Technology},
  publisher = {IOP Publishing},
  author = {Zhang,  Zewen and Paredes,  Roger and Sundar,  Bhuvanesh and Quiroga,  David and Kyrillidis,  Anastasios and Duenas-Osorio,  Leonardo and Pagano,  Guido and Hazzard,  Kaden R A},
  year = {2024},
  month = Nov,
  pages = {015022}
}

@article{XieGroverMixer2025,
  title = {Performance upper bound of a Grover-mixer quantum alternating operator ansatz},
  author = {Xie, Ningyi and Xu, Jiahua and Chen, Tiejin and Lee, Xinwei and Saito, Yoshiyuki and Asai, Nobuyoshi and Cai, Dongsheng},
  journal = {Phys. Rev. A},
  volume = {111},
  issue = {1},
  pages = {012401},
  numpages = {15},
  year = {2025},
  month = {Jan},
  publisher = {American Physical Society}
}

@misc{Brassard_2002,
   title={Quantum amplitude amplification and estimation},
   journal={Quantum Computation and Information},
   publisher={American Mathematical Society},
   author={Brassard, Gilles and Høyer, Peter and Mosca, Michele and Tapp, Alain},
   year={2002},
   pages={53–74} }

\end{document}